\documentclass[12pt]{article}
\usepackage{amsmath}
\usepackage{amsmath,bm}
\usepackage{url}
\usepackage{tikz}
\usetikzlibrary{shapes.geometric}
\usetikzlibrary{arrows}
\usepackage{amsmath}
\usepackage{amssymb}
\usepackage{hyperref}
\usepackage{rotating}
\usepackage{physics}
\usepackage{epsfig}
\usepackage{graphicx}
\usepackage{color}
\usepackage{cite}
\usepackage{subfig}
\usepackage[font=small,labelfont=bf]{caption}

\newcommand{\beq}{\begin{equation}}
\newcommand{\eeq}{\end{equation}}
\newcommand{\ga}{\lower.7ex\hbox{$\;\stackrel{\textstyle>}{\sim}\;$}}
\newcommand{\la}{\lower.7ex\hbox{$\;\stackrel{\textstyle<}{\sim}\;$}}

\hypersetup{
    colorlinks = true,
    citecolor = {blue},
    linkcolor = {blue},
    urlcolor = {blue},
}

\begin{document}

\def\jcap{\ref@jnl{J. Cosmology Astropart. Phys.}}
                % Journal of Cosmology and Astroparticle Physics

\begin{flushright}
{\tt KCL-PH-TH/2019-62}, {\tt CERN-TH-2019-119}  \\
{\tt ACT-04-19, MI-TH-1928} \\
{\tt UMN-TH-3829/19, FTPI-MINN-19/20} \\
\end{flushright}

\vspace{0.2cm}
\begin{center}
{\bf {\large From Minkowski to de Sitter in Multifield No-Scale Models}}

\end{center}
\vspace{0.1cm}

\begin{center}{%\large
{\bf John~Ellis}$^{a}$,
{\bf Balakrishnan~Nagaraj}$^{b}$,
{\bf Dimitri~V.~Nanopoulos}$^{b,c}$, \\[0.1cm]
{\bf Keith~A.~Olive}$^{d}$ and
{\bf Sarunas~Verner}$^{d}$}
\end{center}

\begin{center}
{\em $^a$Theoretical Particle Physics and Cosmology Group, Department of
  Physics, King's~College~London, London WC2R 2LS, United Kingdom;\\
Theoretical Physics Department, CERN, CH-1211 Geneva 23,
  Switzerland;\\
  National Institute of Chemical Physics \& Biophysics, R{\" a}vala 10, 10143 Tallinn, Estonia}\\[0.2cm]
{\em $^b$George P. and Cynthia W. Mitchell Institute for Fundamental
 Physics and Astronomy, Texas A\&M University, College Station, TX
 77843, USA};\\[0.2cm]
{\em $^c$Astroparticle Physics Group, Houston Advanced Research Center (HARC),
 \\ Mitchell Campus, Woodlands, TX 77381, USA;\\ 
Academy of Athens, Division of Natural Sciences,
Athens 10679, Greece}\\[0.2cm]
{\em $^d$William I. Fine Theoretical Physics Institute, School of
 Physics and Astronomy, University of Minnesota, Minneapolis, MN 55455,
 USA}
 
 \end{center}

\vspace{0.1cm}
\centerline{\bf ABSTRACT}
\vspace{0.1cm}

{\small We show the uniqueness of superpotentials leading to Minkowski vacua of
single-field no-scale supergravity models, and the construction of dS/AdS solutions using pairs of these single-field Minkowski superpotentials.
We then extend the construction to two- and multifield no-scale supergravity models, providing also a geometrical interpretation.
We also consider scenarios with additional twisted or untwisted moduli fields, and discuss how inflationary models can be
constructed in this framework.
 } 

\vspace{0.2in}

\begin{flushleft}
{July} 2019
\end{flushleft}
\medskip
\noindent

\newpage

\section{Introduction}

We inhabit a universe with small but non-vanishing vacuum energy that is increasingly well described by a de Sitter geometry
that is almost Minkowski at sub-cosmological scales~\cite{PDG}. Moreover, it is popular to hypothesize that the early universe underwent
a period of near-exponential expansion, called inflation~\cite{reviews}, that might correspond to a near-de Sitter (dS) geometry. These
observations motivate the construction of models that accommodate dS and Minkowski spaces, and may be used to
explore transitions between them.

We expect that physics below the Planck scale is approximately supersymmetric~\cite{Nilles:1983ge,Haber:1984rc}, in which case the appropriate theoretical
framework for studying such cosmological issues is supersymmetry~\cite{cries}, more specifically $\mathcal{N} = 1$ supergravity
in order to accommodate chiral matter fields and general relativity. Generic supergravity models are well known to possess
anti-de Sitter (AdS) vacua and have effective potentials that are far from flat, the `$\eta$-problem' \cite{eta}. However, there is one class
of supergravity models that avoid these problems, namely no-scale supergravity~\cite{no-scale,Ellis:1983sf,LN},
which can accommodate flat potentials that may have vanishing
energy density, corresponding to Minkowski vacua, or have constant positive energy densities, corresponding to dS vacua \cite{EKN1,ENNO}.

Another reason for favouring no-scale supergravity is that it emerges as the natural framework for the low-energy effective field 
theory derived from strings~\cite{Witten}. This was first shown in the context of a simplified model of compactification with a single volume
modulus, but this first example has been extended to multifield models, including compactifications
with three complex K\"ahler moduli and a complex
coupling modulus, as well as some number of complex structure moduli~\cite{FK}.

Several issues then arise within this broader theoretical context. How unique are no-scale supergravity models with Minkowski or de Sitter solutions?
What are the relationships between them? Can they be given simple geometrical interpretations?
How may constructions with a single complex modulus field be generalized to two- or multifield supergravity models?
Can the de Sitter models be used to construct inflationary models predicting perturbations that are consistent with observations,
e.g., resembling the successful~\cite{planck18,rlimit} predictions of the Starobinsky model \cite{Staro} as in~\cite{ENO6}?
How may the universe evolve from a (near-)de Sitter inflationary state towards the (near-)Minkowski contemporary epoch with its
(small) cosmological constant, a.k.a. dark energy?

Aspects of these questions have been addressed previously in a series of papers by subsets of the present authors. In~\cite{ENNO}, we
constructed dS vacua in two- and multifield models as could occur in string compactifications, discussed the conditions for their
stability, and gave examples with only integer powers of the chiral fields in the superpotential. 
There is a long history of no-scale supergravity models of inflation \cite{GL,KQ,EENOS,otherns}, but only recently
has it been realized that simple forms of the superpotential can yield Starobinsky-like inflation \cite{ENO6,Avatars,eno9,ENOV1,ENOV2,ENOV3,king,
KLno-scale,FKR,FeKR,others,rs,EGNO4,reheating,Moreothers,egnno1,egnno23}. Indeed, there are several forms for the superpotential based on 
two chiral fields \cite{Avatars}. 
In~\cite{ENOV1} we presented
a general discussion of two-field no-scale supergravity models of inflation yielding predictions similar to those of the Starobinsky model,
using the non-compact SU(2,1)/SU(2)$\times$U(1) symmetry to catalogue them in six equivalence classes. In~\cite{ENOV2} we
constructed within this framework a specific minimal SU(2,1)/SU(2)$\times$U(1) no-scale model that incorporates
Starobinsky-like inflation, supersymmetry breaking and dark energy. This construction was generalized in~\cite{ENOV3} to
inflationary models based on generalized no-scale structures with different values of the K\"ahler curvature $R$, as may occur if different
numbers of complex moduli contribute to driving inflation.

In this paper we discuss the uniqueness of superpotentials leading to Minkowski, dS and AdS vacua of
single-field no-scale supergravity models, and how pairs of Minkowski superpotentials can be used to construct
dS/AdS solutions. Expanding on previous work which showed how this construction may be extended to two- and multifield no-scale supergravity models, 
we show how matter fields can be incorporated in a multifield construction of Minkowski, dS and AdS vacua. We also
provide a geometrical visualization of the construction. We also mention how Starobinsky-like inflationary models can be
constructed in this framework, and comment on the inclusion of additional twisted or untwisted moduli fields.

The structure of this paper is as follows. In Section~\ref{moduli} we first review the of structure no-scale supergravity
and previous work within that framework. We then discuss the uniqueness of single-field monomial superpotentials leading to a Minkowski
vacuum and how they can be combined in pairs to yield dS vacua. Section~\ref{multimoduli} shows how these constructions
can be extended to multiple moduli, and introduces a geometrical interpretation. Section~\ref{untwisted} then further extends
these constructions to include untwisted matter fields, and Section~\ref{twisted} considers the case of twisted matter fields.
This is followed in Section~\ref{inflation} by a discussion of inflationary models with either untwisted or twisted matter fields.
Finally, our results are summarized in Section~\ref{summary}.

\section{Vacua Solutions with Moduli Fields}
\label{moduli}
\subsection{No-Scale Supergravity Framework}
We first recall some general properties of no-scale supergravity models, which emerge naturally from generic string compactifications in the low-energy effective limit~\cite{Witten}. The simplest $\mathcal{N} = 1$ no-scale supergravity models were first considered in~\cite{no-scale, Ellis:1983sf} and are characterized by the following K\"ahler potential~\cite{EKN1}:
\begin{equation} \label{kah1}
K \; = \; - \, 3 \, \ln (T + \overline{T}),
\end{equation}
where field $T$ is a complex chiral field that can be identified as the volume modulus field, and $\overline{T}$ is its conjugate field. The minimal no-scale K\"ahler potential (\ref{kah1}) describes a non-compact $\frac{SU(1,1)}{U(1)}$ coset manifold and its higher-dimensional generalizations~\cite{EKN2} will be considered in the following sections. Furthermore, the  K\"ahler curvature of a general  K\"ahler manifold is given by the expression $R_{i \bar{j}} \equiv \partial_i \partial_{\bar{j}} \ln K_{i \bar{j}}$, and the scalar curvature obeys the relation:
\begin{equation}
R \; \equiv \; \frac{R_{i \bar{j}}}{K_{i \bar{j}}},
\end{equation}
where $K_{i \bar{j}}$ is the inverse K\"ahler metric. If we consider the maximally-symmetric $\frac{SU(1,1)}{U(1)}$ K\"ahler manifold~(\ref{kah1}), the K\"ahler curvature reduces to the familiar result $R = \frac{2}{3}$. The 
 K\"ahler potential~(\ref{kah1}) can be modified by introducing a curvature parameter $\alpha$:
\begin{equation}\label{kah2}
K \; = \; - \, 3 \, \alpha \, \ln (T + \overline{T} ),
\end{equation}
which also parametrizes a non-compact $\frac{SU(1, 1)}{U(1)}$ coset manifold, but with a positive constant curvature $R = \frac{2}{3 \alpha}$ if we assume that $\alpha > 0$. This unique structure was first discussed in~{\cite{EKN1}}, and similar models were studied in~{\cite{alpha1, rs, alpha3}}, where they were termed $\alpha$-attractors. 

To account for interactions, the K\"ahler potential is extended by including a superpotential $W$:
\begin{equation}
G \equiv K + \ln W + \ln \overline{W} \, ,
\end{equation}
yielding the effective scalar potential:
\begin{equation}\label{effpot}
V = e^G \left[ \pdv{G}{\Phi_i} K_{i \bar{j}} \pdv{G}{\bar{\Phi}_{\bar{j}}} - 3 \right] \, ,
\end{equation}
where the fields $\Phi_i$ are complex scalar fields, $\bar{\Phi}_{\bar{i}}$ are their conjugate fields, and $K_{i \bar{j}}$ is the inverse K\"ahler metric. 
 For more on $N=1$ supergravity models, see \cite{Nilles:1983ge}.

\subsection{Review of Earlier Work}

As was shown in~{\cite{EKN1,rs,eno9}, one can consider combining cubic and constant superpotential terms to acquire a de Sitter vacuum solution. 
Choosing the following superpotential form:
\begin{equation}\label{ds1}
W = 1 - T^3,
\end{equation}
together with the K\"ahler potential~(\ref{kah1}), and imposing the condition $T = \overline{T}$, the effective scalar potential~(\ref{effpot}) yields a de Sitter vacuum solution $V = \frac{3}{2}$. However, the superpotential~(\ref{ds1}) leads to an unstable vacuum solution,
since the mass-squared of the imaginary component of the scalar field is negative: $m_{Im\,T}^2 = -2$.  As we discuss in more detail below,
the problem of instabilities can be addressed by adding a quartic term to the K\"ahler potential \cite{EKN3,Avatars,ENOV3}.

A detailed analysis of the general de Sitter vacua constructions for multi-moduli models was conducted in~\cite{ENNO} and for convenience, we recall some of the key results. 
The Minkowski vacua solutions for a single complex chiral field $T$ were found by considering the K\"ahler potential~(\ref{kah2}) with a monomial superpotential of the following form:
\begin{equation}\label{mink1}
W = \lambda \cdot T^{n_{\pm}} \, ,
\end{equation}
where $n_{\pm}$ are two possible solutions given by:
\begin{equation}\label{npm}
n_{\pm} = \frac{3}{2} \left( \alpha \pm \sqrt{\alpha} \right) \, .
\end{equation}
Along the real $T$ direction, $V = 0$.
The scalar mass-squared in the imaginary direction is:
\begin{equation}\label{massim1}
m_{Im\,T}^2 = 2^{2 - 3 \alpha} \cdot \lambda^2 \cdot \frac{\left( \alpha - 1 \right)}{\alpha} \cdot T^{\pm 3\sqrt{\alpha}} \, ,
\end{equation}
where the choice $T^{\pm 3\sqrt{\alpha}}$ corresponds to the two possible solutions $n_{\pm}$~(\ref{npm}). 
As can be seen from~(\ref{massim1}), in order to obtain a stable Minkowski vacuum solution, the stability condition $\alpha \geq 1$ has to be satisfied. For cases when $0 < \alpha < 1$, quartic stabilization terms in the imaginary direction must be introduced in the K\"ahler potential~(\ref{kah2}).

As was shown in~\cite{EKN1,ENNO,ENOV3}, de Sitter vacua solutions can be obtained from the K\"ahler potential~(\ref{kah2}) by choosing a superpotential of the form:
\begin{equation}\label{ds22}
W =  \lambda_1 \, T^{n_{-}} - \lambda_2 \, T^{n_{+}}  \, ,
\end{equation}
where $n_{\pm}$ is given by~(\ref{npm}). In this case, along the real $T$ direction the effective scalar potential~(\ref{effpot}) becomes:
\begin{equation}
V = 3 \cdot 2^{2 - 3 \alpha} \cdot \lambda_1 \, \lambda_2.
\end{equation}
One of the most fascinating features of the de Sitter vacua construction~(\ref{ds22}) is that it is obtained by combining two distinct Minkowski vacua solutions~(\ref{mink1}). In the next sections, we will show that there is a deeper connection between dS/AdS and Minkowski vacua solutions and that this relation is not accidental.

Superpotential classes yielding constant scalar potentials
were first considered in~\cite{EKN1}, namely:
\begin{eqnarray}
1) \qquad W & = & \lambda \qquad {\rm with} \qquad \alpha = 1 \, , \label{ekn1} \\
2) \qquad W & = & \lambda \, T^{3 \alpha/2} \, , \label{ekn2} \\
3) \qquad  W & = & \lambda \, T^{3\alpha/2} (T^{3 \sqrt{\alpha}/2} - T^{-3 \sqrt{\alpha}/2})\, .  \label{ekn3}
\end{eqnarray}
Comparing solution 1) to~(\ref{ds22}), we see that it can be recovered by setting $\alpha = 1$, 
$\lambda_1 = \lambda$ and $\lambda_2 = 0$. Because $\lambda_2$ is chosen to be zero, 
we find a Minkowski vacuum: $V = 0$.
Solution 3) is identical to (\ref{ds22}) with $\lambda_1 = \lambda_2 =  - \lambda$.
Solution 2)  can also be obtained from (\ref{ds22}) with the aid of a 
K\"ahler transformation:
\begin{equation}\label{kahtrans}
K \rightarrow K + f(T) + \bar{f} (\overline{T})
\end{equation}
and
\begin{equation}
W(T) \longrightarrow \widetilde{W} (T) = e^{-f(T)} W(T),
\label{suptrans}
\end{equation}
with
\begin{equation}\label{func1}
f(T) = \ln (\frac{1 + T^{\, 3 \sqrt{\alpha}}}{2T^{3\sqrt{\alpha}/2}}) \, ,
\end{equation}
and applying the transformation laws~(\ref{kahtrans}),~({\ref{suptrans}}) with~(\ref{func1}) and $\lambda_1= -\lambda_2 = \lambda/2$, 
we recover solution 2) which is in fact an AdS vacuum solution $V = -(3/2^{3\alpha}) \, \lambda^2$, which is always negative. 

While the scalar potential is flat in the real direction, 
the scalar mass-squared of the imaginary component is given by:
\begin{equation}\label{massim2}
m_{Im\,T}^2 = \frac{2^{2 - 3 \alpha}  \left[\lambda_1^2 (\alpha - 1) T^{-3 \sqrt{\alpha}} - 2 \lambda_1 \lambda_2 (\alpha +1) + \lambda_2^2(\alpha - 1) T^{3 \sqrt{\alpha}} \right]}{\alpha} \, ,
\end{equation}
For $\alpha > 0$, in the absence of stabilization terms, there are always some field values for
which the instability in the imaginary direction persists. 
The problem of instability can be remedied by modifying the 
K\"ahler potential~(\ref{kah2}) and introducing quartic stabilization terms in the imaginary direction  \cite{EKN3,Avatars,ENOV3}:
\begin{equation}
K = -3 \, \alpha \ln \left(T + \overline{T} + \beta \, (T - \overline{T})^4 \right),
\label{kahstab}
\end{equation}
with $\beta > 0$. The newly-introduced quartic stabilization term does not alter the potential in the real direction,
while it stabilizes the mass of the imaginary component~(\ref{massim2}) so that:
\begin{equation}\label{massim3}
m_{Im_{\,T}}^2 =  \frac{2^{2 - 3 \alpha} \left[ \lambda_{1}^2 (\alpha -1 + 96 \beta T^3) T^{-3 \sqrt{\alpha }}-2 (\alpha +1 - 96 \beta T^3) \lambda_{1} \lambda_{2} + \lambda_{2}^2 (\alpha -1 + 96 \beta T^3) T^{3 \sqrt{\alpha }} \right]}{\alpha} \, .
\end{equation}

\subsection{Uniqueness of Vacua Solutions}

By solving an inhomogeneous differential equation, we now show that the monomial Minkowski superpotential solutions~(\ref{mink1}) are the only possible unique solutions that yield $V = 0$, while the combination of two distinct Minkowski solutions (\ref{ds22}) yield dS/AdS vacuum solutions. 

We consider a general superpotential expression $W(T)$, which is a function of volume modulus $T$ only, and solve the general homogeneous differential equation, which is equivalent to finding Minkowski vacuum solutions. As before, we assume that the VEV of the imaginary component $\langle Im\,T \rangle = 0$, so that $T = \overline{T}$ and $W(T) = \overline{W} (\bar{T})$. Using the K\"ahler potential (\ref{kah2}) and the effective scalar potential (\ref{effpot}), we find
\begin{equation}\label{dif1}
V = \left(2T \right)^{-3 \alpha} \cdot \left[ \frac{(3 \alpha W - 2 T W')^2}{3 \alpha} - 3W^2 \right] \, ,
\end{equation}
where $W \equiv W(T)$ and $W' \equiv \frac{dW(T)}{dT}$. In order to find Minkowski vacuum solutions, we set Eq.~(\ref{dif1}) to zero:
\begin{equation}\label{dif2}
 \frac{(3 \alpha W - 2 T W')^2}{3 \alpha} - 3W^2  = 0.
\end{equation}
Solving the homogeneous differential equation~(\ref{dif2}), we obtain two distinct Minkowski solutions:
\begin{equation}
W = \lambda_i \cdot T^{\frac{3}{2} \left(\alpha \pm \sqrt{\alpha} \right)},
\end{equation}
where $\lambda_i$ is an arbitrary constant. To find the dS/AdS vacuum solutions, we set the differential equation~(\ref{dif1}) equal to a constant and solve the following inhomogeneous equation:
\begin{equation}\label{dif3}
 \frac{(3 \alpha W - 2 T W')^2}{3 \alpha} - 3W^2= \Lambda \cdot \left(2T \right)^{3 \alpha},
\end{equation}
where $\Lambda$ is an arbitrary constant. We look for a particular superpotential solution to the inhomogeneous equation~(\ref{dif3}) of the following form:
\begin{equation}\label{dif4}
W = \lambda_1 \cdot T^{\frac{3}{2} \left(\alpha \pm \sqrt{\alpha} \right)} - \lambda_2 \cdot T^{\, m} \, .
\end{equation}
Inserting the expressions~(\ref{dif4}) into (\ref{dif3}), we find that $m = n_\mp = \frac{3}{2} \left(\alpha \mp \sqrt{\alpha} \right)$ is a particular solution 
of the inhomogeneous differential equation and the general solution
has the following form:
\begin{equation}	
W =  \lambda_1 \cdot T^{n_{\pm}} - \lambda_2 \cdot T^{\, n_{\mp}},~\text{with} \quad V = \Lambda \, ,
\end{equation}
where we have defined the constant $\Lambda = 3 \cdot 2^{2 - 3 \alpha} \cdot \lambda_1 \, \lambda_2 $. Thus, 
we have constructed the unique combination of two Minkowski solutions that yields dS/AdS solutions~\footnote{We note that these solutions correspond to flat directions in the real field direction.  It is possible and
relatively straightforward to construct minima with non-zero vacuum energy. In particular, it is well known
that supergravity models with unbroken supersymmetry generally lead to AdS vacua.}.

\subsection{Generalized Solutions and Vacuum Stability}

Before concluding this section, we introduce a formalism with which
the construction of Minkowski-dS-AdS solutions can be generalized and applied to more complicated K\"ahler manifolds. Let us write:
\begin{equation}\label{kahv}
K = -3 \, \alpha \ln(\mathcal{V}) \, ,
\end{equation}
where $\mathcal{V}$ is the argument inside the logarithm.
For the simplest minimal no-scale $\frac{SU(1,1)}{U(1)}$ supergravity case with a single volume modulus field $T$, we have
$\mathcal{V} \equiv T + \overline{T}$.
As before, and in all the cases that we consider, we assume that the VEV of the imaginary part of the complex field is fixed to zero: $\langle Im\,T \rangle = 0$, which can always be achieved by introducing quartic stabilization terms in Eq.~(\ref{kahv}).

For the single field case, the effective scalar potential~(\ref{effpot}) becomes:
\begin{equation}\label{effpotv}
V = \frac{\hat{V}}{\mathcal{V}^{3 \alpha}},~\text{with} \quad
\hat{V} = \frac{|\mathcal{V} \cdot \overline{W}_{\overline{T}} - 3 \alpha W|^2}{3 \alpha} - 3 |W|^2.
\end{equation}
In the real direction, where $T = \overline{T}$, we define:

\begin{equation}
\mathcal{V} \longrightarrow \xi ,
\end{equation}
so that the argument inside the logarithm becomes $\xi = 2 T$.

From our previous discussion, we already know which superpotential forms reduce to Minkowski solutions. 
We introduce the following notation, which will be used for all our K\"ahler coset manifolds~\footnote{Note that we are using a trick in our definition of the 
superpotential. Strictly speaking, $\xi$ is defined by the argument of the $\log$ in $K$ when all fields are taken as real. However in the superpotential we are assuming that $\xi$ is a function of (complex) superfields and ignore the restriction to real fields.} :	
\begin{equation}\label{mink22}
W_{M} \; \equiv \; \lambda \cdot \xi^{n_{\pm}} \, ,~\text{with} \quad V=0 \, ,
\end{equation}
where, as usual, the two possible choices $n_{\pm}$ are given by Eq.~(\ref{npm}). Note that, for this construction to work, we must impose the constraint $\xi > 0$ and the positive curvature condition $\alpha > 0$, which are necessary features of the
no-scale structure~\footnote{Note also that the definition of $\lambda$ here differs from that in Eq. (\ref{mink1}) by a constant factor
of $2^{n_\pm}$.}.

With this redefinition,  the scalar mass-squared in the imaginary field direction given in Eq. (\ref{massim1}) becomes:
\begin{equation}\label{massmink1}
m_{Im\,T}^{2} = \frac{4 \, \lambda^2 (\alpha -1) \xi^{\pm 3 \sqrt{\alpha }}}{\alpha } \, ,
\end{equation}
where the sign depends on the choice of the Minkowski vacuum solution in~(\ref{mink22}). We will later show that the same Minkowski mass expression~(\ref{massmink1}) holds for any K\"ahler potential form, and hence that the solution is stable when $\alpha \geq 1$.  When $0 < \alpha < 1$, Minkowski vacuum solutions become unstable and we must introduce the quartic stabilization terms in the imaginary direction.

Similarly, dS/AdS vacua solutions are constructed by combining two different Minkowski solutions~(\ref{mink22}),
\begin{equation}\label{ds3}
W_{dS/AdS} = \lambda_{1} \cdot \xi^{n_{-}} - \lambda_{2} \cdot \xi^{n_{+}} \, ,
\end{equation}
and we call such constructions Minkowski pairs.
The dS/AdS vacuum solution~(\ref{ds3}) yields an effective scalar potential~(\ref{effpot}):
\begin{equation}\label{potds}
V = 12 \,\lambda_{1} \, \lambda_{2} \, ,
\end{equation}
 which allows three different types of vacua:
\begin{itemize}
\item{de Sitter vacuum solutions when $\lambda_1$ and $\lambda_2$ are $\ne 0$ and have the same sign.}
\item{anti-de Sitter vacuum solutions when $\lambda_1$ and $\lambda_2$ are $\ne 0$ and have opposite signs.}
\item{Minkowski vacuum solutions when either $\lambda_1$ or $\lambda_2$ is set to zero.}
\end{itemize}

The generalization of the scalar mass in the imaginary direction $m_{Im \, T}^2$ in equation~(\ref{massim2}) is given by:
\begin{equation}\label{massds1}
m_{Im{\,T}}^2 =  \frac{4 \left[ \lambda_{1}^2 (\alpha -1) \xi ^{-3 \sqrt{\alpha }}-2 (\alpha +1) \lambda_{1} \lambda_{2} + \lambda_{2}^2 (\alpha -1)  \xi ^{3 \sqrt{\alpha }} \right]}{\alpha} \, ,
\end{equation}
which should always be positive, $m_{Im\,T}^2 \geq 0$ for stability in the imaginary field direction. 
Recalling that dS vacuum solutions are acquired when $\lambda_{1}$ and $\lambda_2$ have the same sign, we introduce the ratio coefficient $\gamma = \lambda_1/\lambda_2$, which must always be positive.

To visualize this condition, we plot in Fig.~\ref{color} the $(\alpha, \gamma)$ plane with $T$ on the vertical axis, and
the size of $\log \left( m_{Im{\,T}}^2 / 4 \, \lambda_2^2  \right)$ indicated by color coding. The boundary of the colored region corresponds to the critical value of $m_{Im{\,T}}^2 / 4 \, \lambda_2^2 = 0$, 
and it indicates when $m_{Im{\,T}}^2$ becomes unstable. Interestingly, the same general expression~(\ref{massds1}) 
holds also for more complicated forms of $\xi$.  It is important to note that Fig.~\ref{color} shows two colored regions which are separated by a gap, and indicates that the dS vacuum becomes unstable in the imaginary direction for certain values of $T$ and $\alpha$. 

\begin{figure}[h!]
\centering
\includegraphics[scale=0.6]{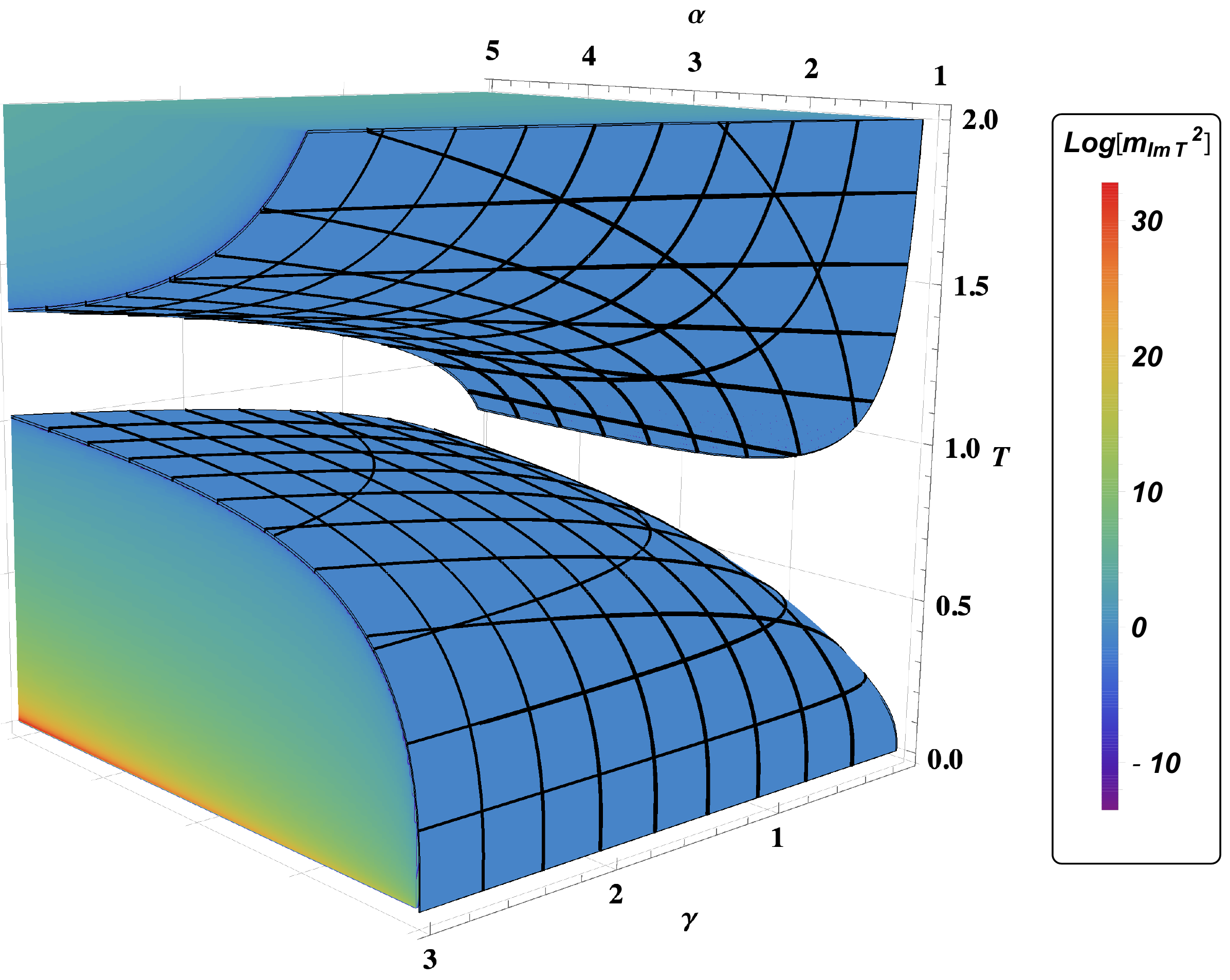}
\caption{\it Illustration of the value of the expression~({\ref{massds1}}) as a function of $(\alpha, \gamma, T)$,
as shown by the color coding for $\log \left( m_{Im{\,T}}^2 / 4 \, \lambda_2^2  \right)$ on the right-hand side.}
\label{color}
\end{figure}

To understand the occurrence of the dS vacuum instability, we consider two specific cases with different values of $\alpha$, where for illustrative purposes we choose $\lambda_1 = \lambda_2 = 1$, and we use the field parametrization $T = (x + i y)/\sqrt{2}$. The effective scalar potential is plotted in the left panel of Fig.~\ref{dsplot1} for $\alpha = 1$, which is characteristic of solutions with $\alpha \leq  1$. We see that dS vacuum solutions are always unstable in the imaginary field direction, so these solutions must be stabilized. In the right panel of
Fig.~\ref{dsplot1} we show the scalar potential with $\alpha = 3$, which is characteristic of solutions with  $\alpha > 1$. Here, we see that vacuum solutions might fall into an AdS vacuum, which corresponds to the gap region shown in Fig.~\ref{color}. In both cases, the potential is completely 
flat along the line $y = 0$ corresponding to the dS solution up to the point where $x=0$ (the potential is not defined at $x \le 0$). 

\begin{figure}%
    \centering
    \subfloat[$\alpha = 1$]{{\includegraphics[scale=0.42]{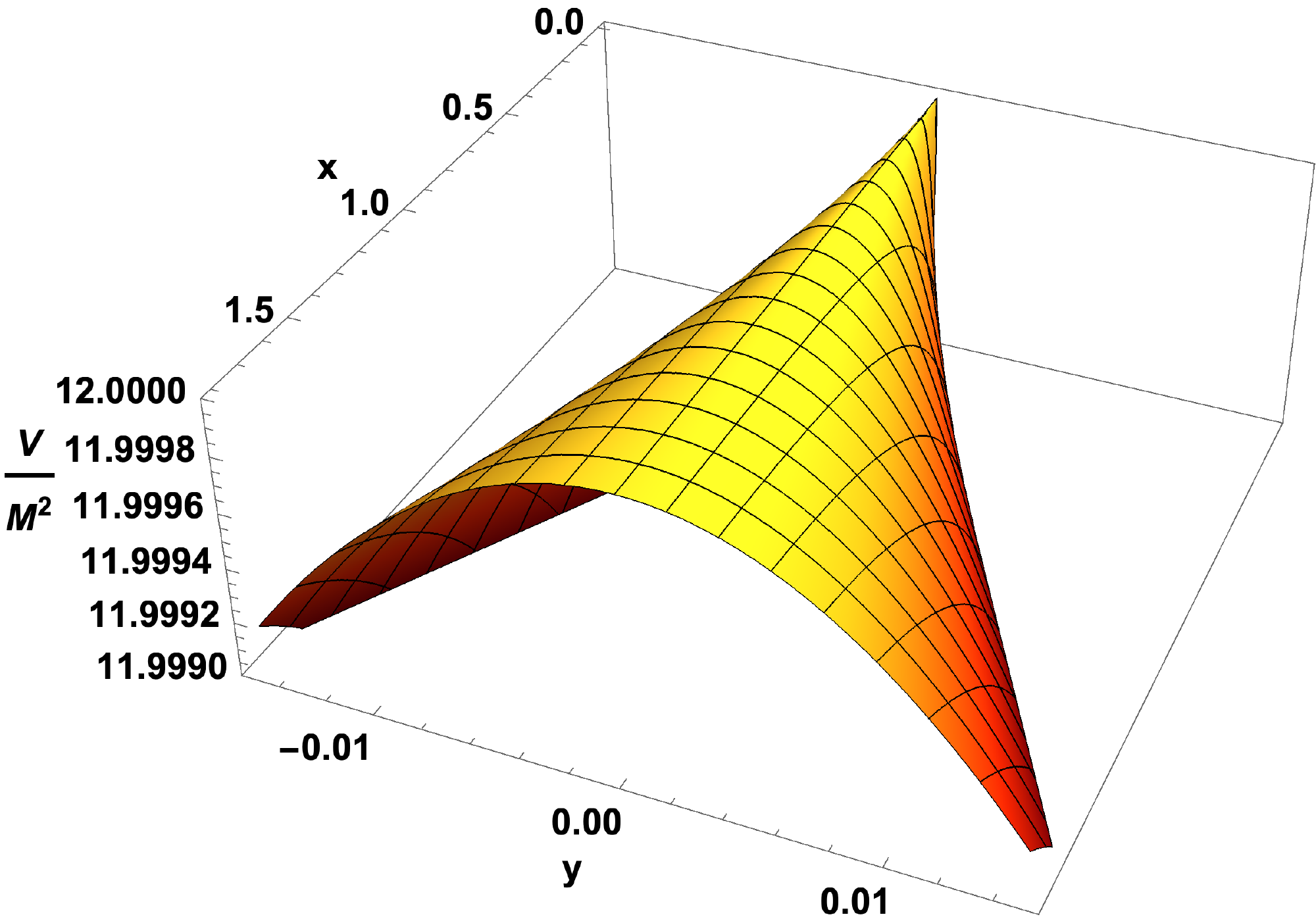} }}%
  %  \qquad
    \subfloat[$\alpha = 3$]{{\includegraphics[scale=0.42]{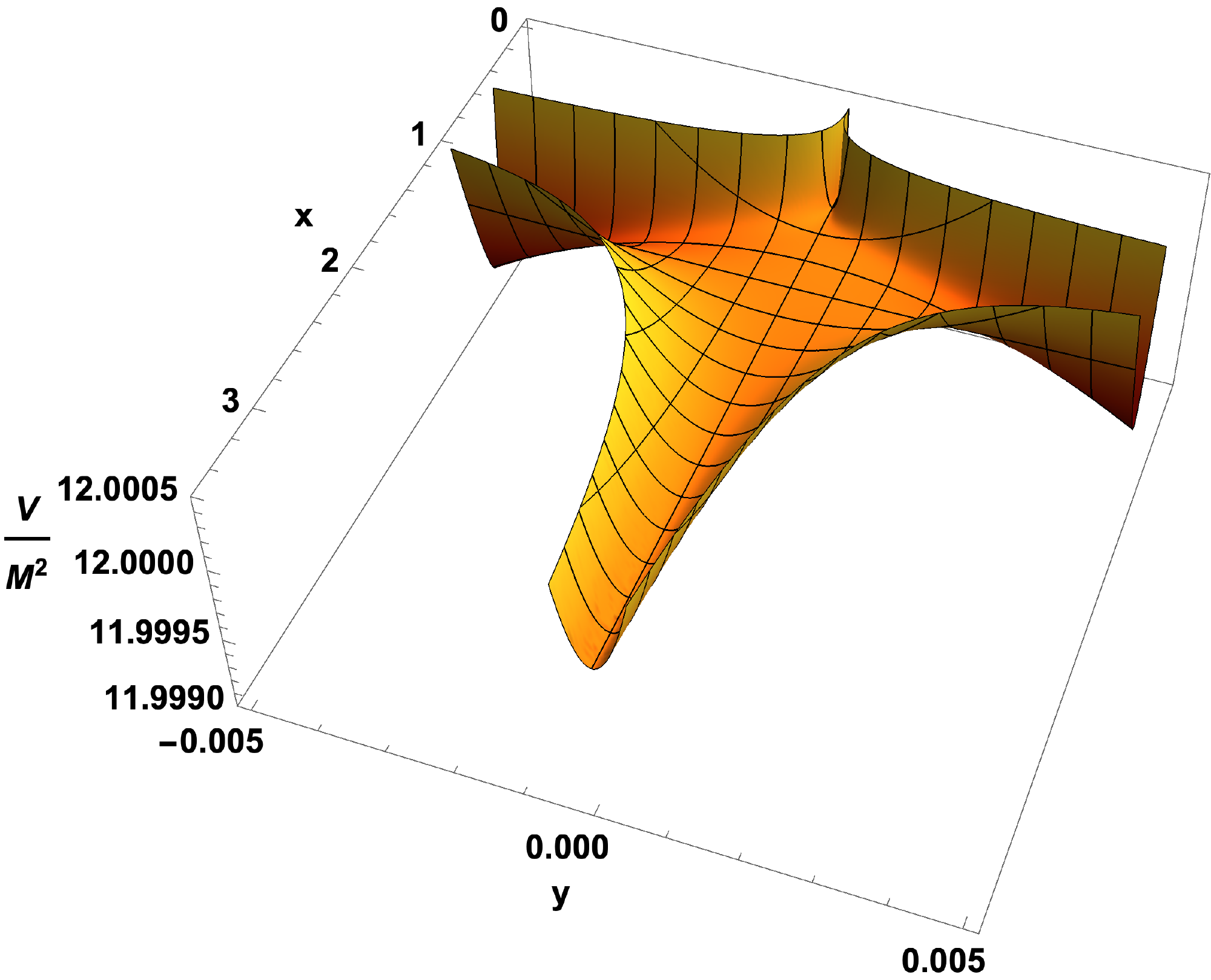} }}%
    \caption{\it The effective scalar potential $V(x, y)$ without quartic stabilization terms in the imaginary direction ($\beta = 0$), for the cases $\alpha = 1$ (left panel) and $\alpha = 3$ (right panel).}%
    \label{dsplot1}%
\end{figure}

To address the stability issue, we consider the modified K\"ahler potential~(\ref{kahstab}), where if we compare it to the general K\"ahler potential (\ref{kahv}), we see that in the real direction the argument inside the logarithm remains unchanged, with $\xi = 2 T$.

The generalization of the mass squared 
in Eq. (\ref{massds1}) is:
\begin{equation}\label{massds3}
m_{Im{\,T}}^2 =  \frac{4 \left[ \lambda_{1}^2 (\alpha -1 + 12 \beta \cdot \xi^3) \xi ^{-3 \sqrt{\alpha }}-2 (\alpha +1 - 12 \beta \cdot \xi^3) \lambda_{1} \lambda_{2} + \lambda_{2}^2 (\alpha -1 + 12 \beta \cdot \xi^3)  \xi ^{3 \sqrt{\alpha }} \right]}{\alpha} \, ,
\end{equation}
where it can readily be seen from the numerator of~(\ref{massds3}) that, 
by choosing a value of $\beta$ that is large enough, we can always make the imaginary field direction 
stable~\footnote{A similar expression when $\gamma = 1$ can be found
in \cite{ENNO}.}.  We plot in Fig.~\ref{dsplot2} the unstable cases considered previously 
with $\alpha = 1$ and $\alpha = 3$, which have been each stabilized with the choice $\beta = 2$. 
Once again, the potential along $y=0$ is flat. 

\begin{figure}%
    \centering
    \subfloat[$\alpha = 1$]{{\includegraphics[scale=0.42]{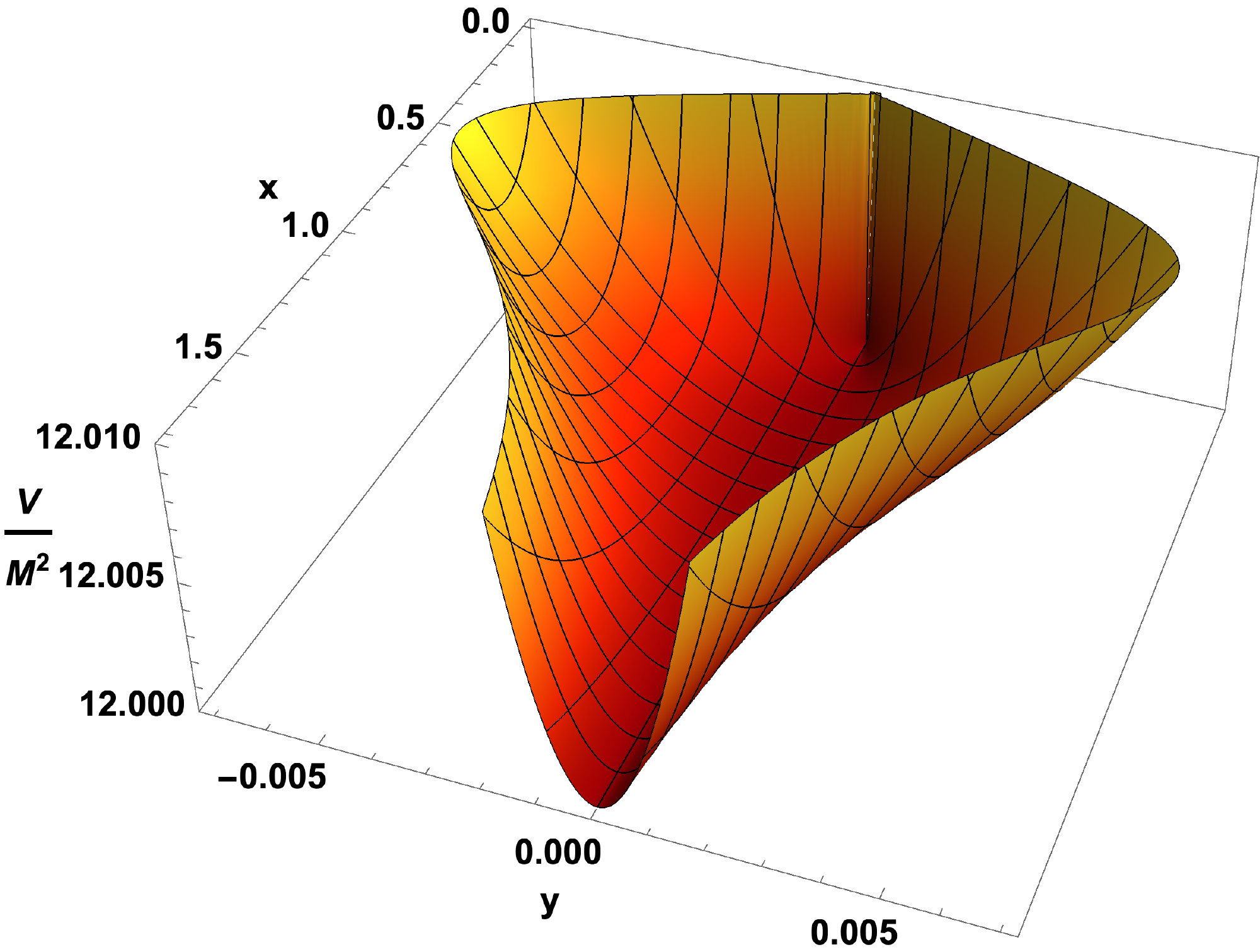} }}%
%    \qquad
    \subfloat[$\alpha = 3$]{{\includegraphics[scale=0.42]{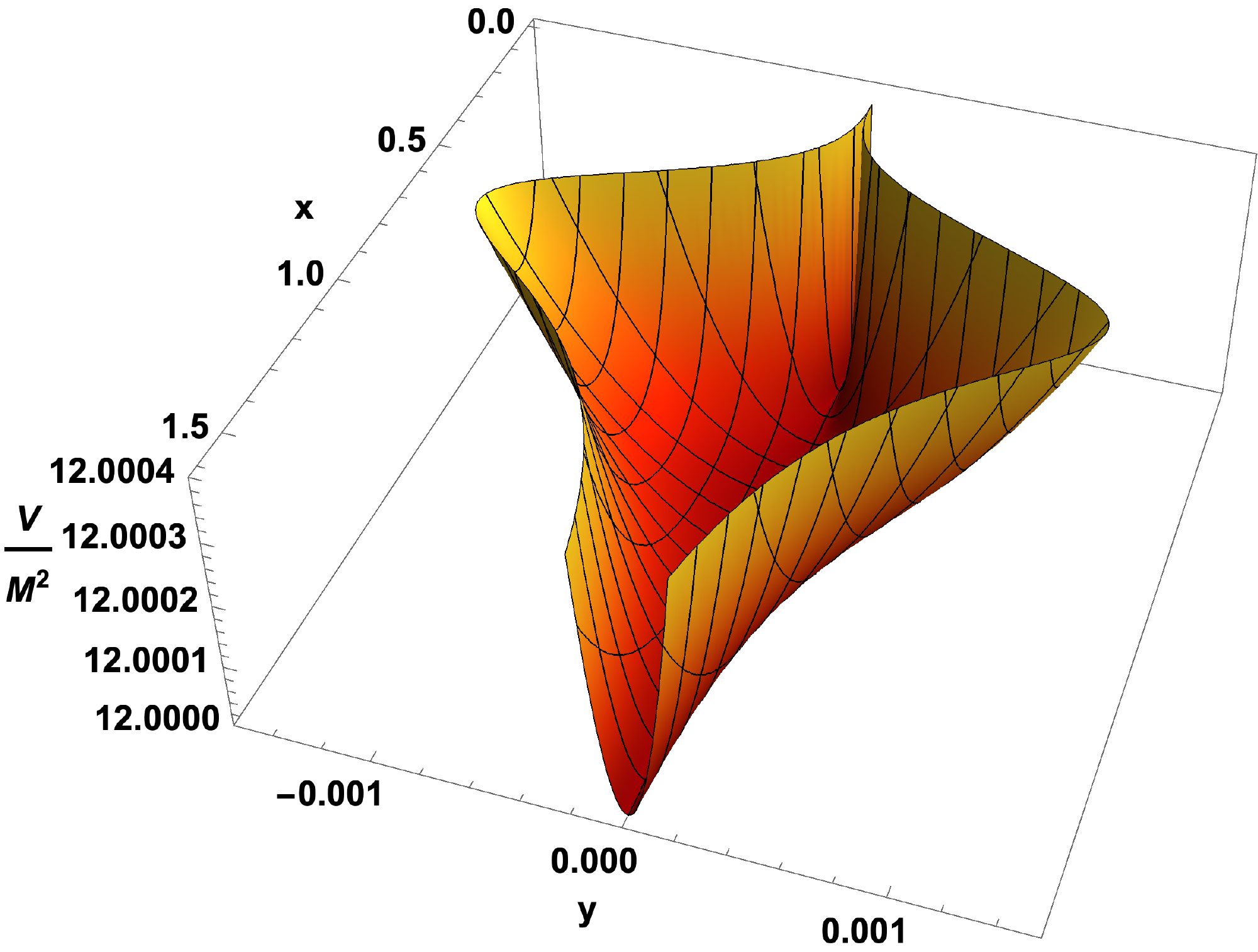} }}%
    \caption{\it The effective scalar potential $V(x, y)$ for the two values $\alpha = 1$ (left panel)
    and $\alpha = 3$ right panel, now stabilized by quartic terms in the imaginary direction with $\beta = 2$.}%
    \label{dsplot2}%
\end{figure}

\section{Multi-Moduli Models}
\label{multimoduli}
\subsection{Minkowski Vacuum for Two Moduli}

Our next step is to extend this formulation to the two- and multi-moduli cases.
As before, we first construct the general Minkowski vacuum solutions and then use
Minkowski superpotential pairs to obtain dS/AdS solutions. We begin by considering the following two-field K\"ahler  potential:
\begin{equation}
K \; = \; -3 \, \alpha_1 \ln(\mathcal{V}_1) - 3 \, \alpha_2 \ln (\mathcal{V}_2) \, .
\label{kah3}
\end{equation}
For now, we consider $\mathcal{V}_1 = T_1 + \overline{T}_1$ and $\mathcal{V}_2 = T_2 + \overline{T}_2$. 
Along the real directions, 
$T_1 = \overline{T}_1$ and $T_2 = \overline{T}_2$, we adopt the following notation:
\begin{equation}
\mathcal{V}_1 \rightarrow \xi_1, \qquad \mathcal{V}_2 \rightarrow \xi_2 \, ,
\end{equation}
and we choose the following ansatz that yields Minkowski vacuum solutions:
\begin{equation}
W_{M} = \lambda \cdot \xi_{1}^{n_1} \cdot \xi_{2}^{n_2} \, .
\label{mink2}
\end{equation}
Inserting the superpotential~(\ref{mink2}) into the expression~(\ref{effpot})
for the effective scalar potential, we obtain:
\begin{equation}
V = \lambda^2 \cdot \xi_1^{2 n_1 - 3 \alpha_1} \cdot \xi_2^{2 n_2 - 3 \alpha_2} \cdot \left( \frac{\left(2 n_1-3 \alpha _1\right){}^2}{3 \alpha _1}+\frac{\left(2 n_2-3 \alpha _2\right){}^2}{3 \alpha _2}-3 \right) \, .
\end{equation}
In order to recover Minkowski vacua, we set $V = 0$, which holds when the following expression is satisfied \cite{ENNO}:
\begin{equation}
\frac{\left(2 n_1-3 \alpha _1\right){}^2}{3 \alpha _1}+\frac{\left(2 n_2-3 \alpha _2\right){}^2}{3 \alpha _2} = 3 \, .
\label{circ1}
\end{equation}
For ease of illustration, we introduce the following parametrization:
\begin{equation}
r_1 \equiv \frac{2 n_1 - 3 \alpha_1}{3 \sqrt{\alpha_1}} \, , \qquad 
r_2 \equiv \frac{2 n_2 - 3 \alpha_2}{3 \sqrt{\alpha_2}} \, ,
\label{circ2}
\end{equation}
in terms of which the general expression~(\ref{circ1}) becomes:
\begin{equation}
r_1^2 + r_2^2 = 1 \, .
\label{circ3}
\end{equation}
Solving the constraint~(\ref{circ1}) for $n_{1}$ and $n_{2}$, we find:
\begin{equation}
n_1 = \frac{3}{2} \left( \alpha_1 \pm \sqrt{1 - \frac{(2n_2 - 3 \alpha_2)^2}{9 \alpha_2}} \cdot \sqrt{\alpha_1} \right) \quad~\text{and}~ \quad
n_2 = \frac{3}{2} \left( \alpha_2 \pm \sqrt{1 - \frac{(2n_1 - 3 \alpha_1)^2}{9 \alpha_1}} \cdot \sqrt{\alpha_2} \right) \, ,
\end{equation}
which can be parametrized using~(\ref{circ2}):
\begin{equation}
n_{1} = \frac{3}{2} \left(\alpha_1 + r_1 \sqrt{\alpha_1} \right) \quad~\text{and}~ \quad
n_{2} = \frac{3}{2} \left(\alpha_2 + r_2 \sqrt{\alpha_2} \right) \, ,
\label{npm1}
\end{equation}
where the values of $r_1$ and $r_2$ are constrained by expression~(\ref{circ3}),
and must satisfy the condition $r_i \in \{-1,  1\}$. It can already be seen from these equations 
that the circular parametrization (\ref{circ2}) simplifies our expressions significantly,
and it will be useful in establishing a geometric connection. We must also satisfy the following inequalities:
\begin{equation}
\alpha_i > 0, \quad \text{with}~i = 1, 2 \, .
\end{equation}
We see from~(\ref{npm1}) that we can consider a total of four different sign combinations that
yield $V = 0$. The corresponding expressions for the imaginary masses-squared are given by:
\begin{equation}
m_{Im\,T_i}^2 = \frac{4 \, \lambda^2 \cdot \xi_1^{3 r_1 \sqrt{\alpha_1}} \cdot \xi_2^{3 r_2 \sqrt{\alpha_2}} \left(\alpha_i - r_i^2 \right) }{\alpha_i}, \quad~\text{with}~i=1,2 \, ,
\end{equation}
where stability in the imaginary direction is obtained when the condition $\alpha_i - r_{i}^2 \geq 0$ is satisfied. If we this combine this inequality with the constraint (\ref{circ3}), we obtain another stability condition in terms of the curvature parameters:
\begin{equation}
\alpha_1 + \alpha_2 \geq 1.
\end{equation}

\subsection{Minkowski Pair Formulation for Two Moduli}

Applying the same approach that we used for the case of a single modulus, we now
show how to construct Minkowski pairs for the two-field case 
and recover dS/AdS vacuum solutions with $V = 12 \, \lambda_1 \, \lambda_2$ (as in (\ref{potds})) along the direction where 
all fields are real. 
The general dS/AdS vacuum solutions for the two-field case are given by:
\begin{equation}
W_{dS/AdS} = \lambda_1 \cdot \xi_1^{n_{1}} \cdot \xi_2^{n_{2}} - \lambda_2 \cdot \xi_1^{\bar{n}_{1 }} \cdot \xi_2^{\bar{n}_{2}} \, ,
\label{ds33}
\end{equation}
where we define:
\begin{equation}
\bar{n}_{1} \equiv \frac{3}{2} \left(\alpha_1 + \bar{r}_1 \sqrt{\alpha_1} \right) \quad~\text{and}~ \quad
\bar{n}_{2} \equiv  \frac{3}{2} \left(\alpha_2 + \bar{r}_2 \sqrt{\alpha_2} \right) \, ,
\label{npmc1}
\end{equation}
with the expressions for $n_{1,2}$ being given by Eq.~(\ref{npm1}) and $\bar{r}_i = -r_i$. 
We note that the powers~(\ref{npmc1}) 
describe the antipode of a point lying on the surface of a circle described by the coordinates $(r_1, r_2)$,
and we discuss the geometric interpretation of our models in the next Section. 

The scalar masses recovered from the dS/AdS superpotential~(\ref{ds33}) have complicated expressions
that we do not list here. However, we note that we can always modify the initial K\"ahler potential~(\ref{kah3}) 
by including higher-order corrections in the imaginary direction:
\begin{equation}
K = -3 \, \alpha_1 \ln \left(T_1 + \overline{T}_1 + \beta_1 \left(T_1 - \overline{T}_1 \right)^4 \right) -3 \, \alpha_2 \ln \left(T_2 + \overline{T}_2 + \beta_2 \left(T_2 -  \overline{T}_2 \right)^4 \right),
\end{equation}
where these quartic terms easily remedy the stability problems \cite{ENNO}. If we compare it to the general two-field K\"ahler potential in Eq.~(\ref{kah3}), along the real directions, $T_1 = \overline{T}_1$ and $T_2 = \overline{T}_2$, we recover $\xi_1 = 2T_1$ and $\xi_2 = 2 T_2$. In the next Section we extend this formulation to the $N$-field case.

\subsection{Minkowski Pair Formulation for Multiple Moduli}

We now show how to generalize our formulation and construct successfully the Minkowski pair superpotential 
for cases with $N > 2$ moduli. We first introduce the following K\"ahler potential:
\begin{equation}
K = -3 \sum_{i = 1}^N \, \alpha_i \ln \left(\mathcal{V}_i \right) \, ,
\label{kahmult}
\end{equation}
where $\mathcal{V}_i = T_i + \overline{T}_i$. Next, we impose the condition that all our fields are real, 
therefore $T_i = \overline{T}_i$, which leads to:
\begin{equation}
\mathcal{V}_i \longrightarrow \xi_i, \quad \text{for}~i = 1,2,...,N \, .
\end{equation}
Minkowski vacuum solutions are obtained with the choice:
\begin{equation}
W_{M} = \lambda \cdot \prod_{i =1}^N \xi_i^{n_{i}} \, 
\label{mink3}
\end{equation}
in the general $N$-field case. Inserting the superpotential~(\ref{mink3}) into Eq.~(\ref{effpot}), we find:
\begin{equation}
V = \lambda^2 \cdot \prod_{i = 1}^N \xi_{i}^{2 n_{i} - 3 \alpha_i} \cdot \left( \sum_{i = 1}^N \frac{(2n_i - 3 \alpha_i)^2}{3 \alpha_i} - 3 \, \right)\, ,
\label{mink4}
\end{equation}
and it can be seen from Eq.~(\ref{mink4}) that in order to obtain Minkowski vacuum solutions: $V = 0$, 
we must satisfy the constraint:
\begin{equation}
\sum_{i = 1}^N  \frac{(2n_i - 3 \alpha_i)^2}{3 \alpha_i} = 3 \, .
\label{cond1}
\end{equation}
Once again, we introduce the following parametrization:
\begin{equation}
r_i \equiv \frac{2n_i - 3 \alpha_i}{3 \sqrt{\alpha_i}}, \quad~{\text{for}}~i=1, 2, ..., N \, ,
\label{cond2}
\end{equation}
and combining the equations~(\ref{cond1}) and~(\ref{cond2}) we obtain:
\begin{equation}
\sum_{i = 1}^N r_i^2 = 1 \, .
\label{cond3}
\end{equation}
Therefore, Eq.~(\ref{cond3}) parametrizes the $N$-field Minkowski solutions as
lying on the surface of an $\left(N-1 \right)$-sphere.

Solving Eq.~(\ref{cond2}) for $n_i$, we obtain:
\begin{equation}
n_{i} = \frac{3}{2} \left( \alpha_i + r_i \sqrt{\alpha}_i \right), \quad~{\text{for}}~i=1, 2, ..., N \, ,
\end{equation}
where $r_ i \in \{-1, 1\}$ and $\alpha_i > 0$. For the $N$-moduli case, 
we obtain the following expression for the scalar masses-squared in the imaginary directions:
\begin{equation}
m_{Im\,T_i}^2 = \frac{4 \lambda^2 \left(\alpha_i - r_i^2 \right) \prod_{i = 1}^N \xi_i^{-3 r_i \sqrt{\alpha_i}}}{\alpha_i}, \quad \text{with}~i = 1,2, ..., N \, .
\end{equation}
To obtain a stable solution in the imaginary direction, we must satisfy the condition $\alpha_i - r_i^2 \geq 0$. If we use the constraint of the $(N-1)$-sphere~(\ref{cond3}), we obtain the following stability condition:
\begin{equation}
\sum_{i = 1}^{N} \alpha_i \geq 1.
\end{equation}

Following the procedure described previously, we combine a pair of Minkowski solutions~(\ref{mink3}) and introduce the following dS/AdS superpotential:
\begin{equation}
W_{dS/AdS} = \lambda_1 \cdot \prod_{i=1}^N \xi_i^{n_{i}} - \lambda_2 \cdot \prod_{i=1}^N \xi_i^{\bar{n}_i}  \, , 
\label{mink5}
\end{equation}
where $\bar{n}_i = \frac{3}{2} \left( \alpha_i + \bar{r}_i \sqrt{\alpha}_i \right)$, with $\bar{r}_i = - r_i$. 
This superpotential form also yields the familiar dS/AdS vacuum result $V = 12 \, \lambda_1 \, \lambda_2$.

It proves difficult to perform a detailed stability analysis for $N$-moduli models,
because this would involve finding the eigenvalues of an $N \cross N$ matrix. Nevertheless, 
one can always introduce higher-order corrections in the K\"ahler potential~(\ref{kahmult}):
\begin{equation}
K = -3 \sum_{i = 1}^N \alpha_i \ln \left(T_i + \overline{T}_i + \beta_i \left(T_i - \overline{T}_i  \right)^4 \right),
\label{kah55}
\end{equation}
where the quartic terms stabilize the imaginary directions~\cite{ENNO}. If we compare the multi-moduli K\"ahler potential~(\ref{kahmult}) with~(\ref{kah55}), we see that along the real directions, $T_i = \overline{T}_i$, and we recover $\xi_i = 2 T_i$.

\subsection{Geometric Interpretation}

We now discuss the geometric interpretation of this Minkowski pair formulation.
From equations~(\ref{cond1}-\ref{cond3}) it is clear that our parametrization describes 
Minkowski superpotential solutions~(\ref{mink3}) that lie on the surface of an 
$\left(N-1 \right)$-sphere that is embedded in Euclidean $N$-space. 
We first return to the two-moduli case, in which the Eq.~(\ref{cond3}) reduces to~(\ref{circ3}),
and all Minkowski solutions lie on a circle embedded in 2-dimensional space. We define the radius vector of 
points on a circle $r$ by:
\begin{equation}
r \; = \; \left(r_1, r_2 \right), \quad~\text{with}~r_1^{2} + r_2^{2} = 1.
\label{radius1}
\end{equation}
As expected, equation~(\ref{radius1}) includes 4 possible sign combinations corresponding to 
different quadrants of a circle. To construct successfully a Minkowski pair superpotential that yields 
a dS/AdS vacuum solution, we must combine any chosen point on the circle with its antipodal point, given by the vector:
\begin{equation}
\bar{r} \; = \; -r = -\left(r_1, r_2 \right) \, .
\label{radius2}
\end{equation}
In this way, we can construct an infinite number of distinct Minkowski superpotential pairs by 
considering different point/antipode combinations lying on the surface of a circle. 
The Minkowski pair construction on a circle is illustrated in~Fig.~\ref{circle}. For any value of $\alpha > 0$,
Eq.~(\ref{ds33}) will yield a dS or AdS solution so long as ${n}_i = \frac{3}{2} \left( \alpha_i + {r}_i \sqrt{\alpha}_i \right)$ and $\bar{n}_i = \frac{3}{2} \left( \alpha_i + \bar{r}_i \sqrt{\alpha}_i \right)$.

\begin{figure}[h!]
\begin{tikzpicture}[ scale=5.,cap=round,>=latex]
        % draw the coordinates
        \tikzset{font=\large}
        \draw[->] [line width=0.4mm](-1.5cm,0cm) -- (1.5cm,0cm) node[right,fill=white] {$r_1$};
        \draw[->] [line width=0.4mm](0cm,-1.5cm) -- (0cm,1.5cm) node[above,fill=white] {$r_2$};

        % draw the unit circle
        \draw[line width=0.4mm] (0cm,0cm) circle(1cm);

        \foreach \x in {30, 210} {
                % lines from center to point
                \draw[red][line width=0.4mm] (0cm,0cm) -- (\x:1cm);
                % dots at each point
                \filldraw[red] (\x:1cm) circle(0.5pt);
                % draw each angle in degrees
                \draw (\x:0.6cm) node[fill=white] {$\x^\circ$};
        }
        
\foreach \x in {120, 300} {
                % lines from center to point
                \draw[blue] [line width=0.4mm](0cm,0cm) -- (\x:1cm);
                % dots at each point
                \filldraw[blue] (\x:1cm) circle(0.5pt);
                % draw each angle in degrees
                \draw (\x:0.6cm) node[fill=white] {$\x^\circ$};
        }
        
\foreach \x in {0, 0} {
                % lines from center to point
                \draw[black][line width=0.4mm] (0cm,0cm) -- (\x:0cm);
                % dots at each point
                \filldraw[black] (\x:0cm) circle(0.4pt);
                % draw each angle in degrees
                
        }
        
        \foreach \x/\xtext/\y in {
            % the coordinates for the first quadrant
            30/\frac{\sqrt{3}}{2}/\frac{1}{2},
            120/-\frac{1}{2}/\frac{\sqrt{3}}{2},
            % the coordinates for the third quadrant
            210/-\frac{\sqrt{3}}{2}/-\frac{1}{2},
            300/\frac{1}{2}/-\frac{\sqrt{3}}{2}}
                \draw (\x:1.28cm) node[fill=white] {$\left(\xtext,\y\right)$};

        % draw the horizontal and vertical coordinates
        % the placement is better this way
        \draw (-1.15cm,0cm) node[above=1pt] {$(-1,0)$}
              (1.15cm,0cm)  node[above=1pt] {$(1,0)$}
              (0cm,-1.15cm) node[fill=white] {$(0,-1)$}
              (0cm,1.15cm)  node[fill=white] {$(0,1)$};
    \end{tikzpicture}
    \caption{\it Depiction of Minkowski pairs on a circle. The circle is split into four quadrants and two distinct 
    Minkowski pairs are shown lying in different quadrants. The red dots show a Minkowski pair solution 
    $r = (\sqrt{3}/2 , 1/2)$ and $\bar{r} = (-\sqrt{3}/2 , -1/2)$, which lies in the first and third quadrants of the circle, 
    while the blue dots show a Minkowski pair solution  $r =(-1/2, \sqrt{3}/2)$ and $\bar{r} = (1/2 , -\sqrt{3}/2)$, 
    which lies in the second and fourth quadrants of the circle.}
        \label{circle}
\end{figure}
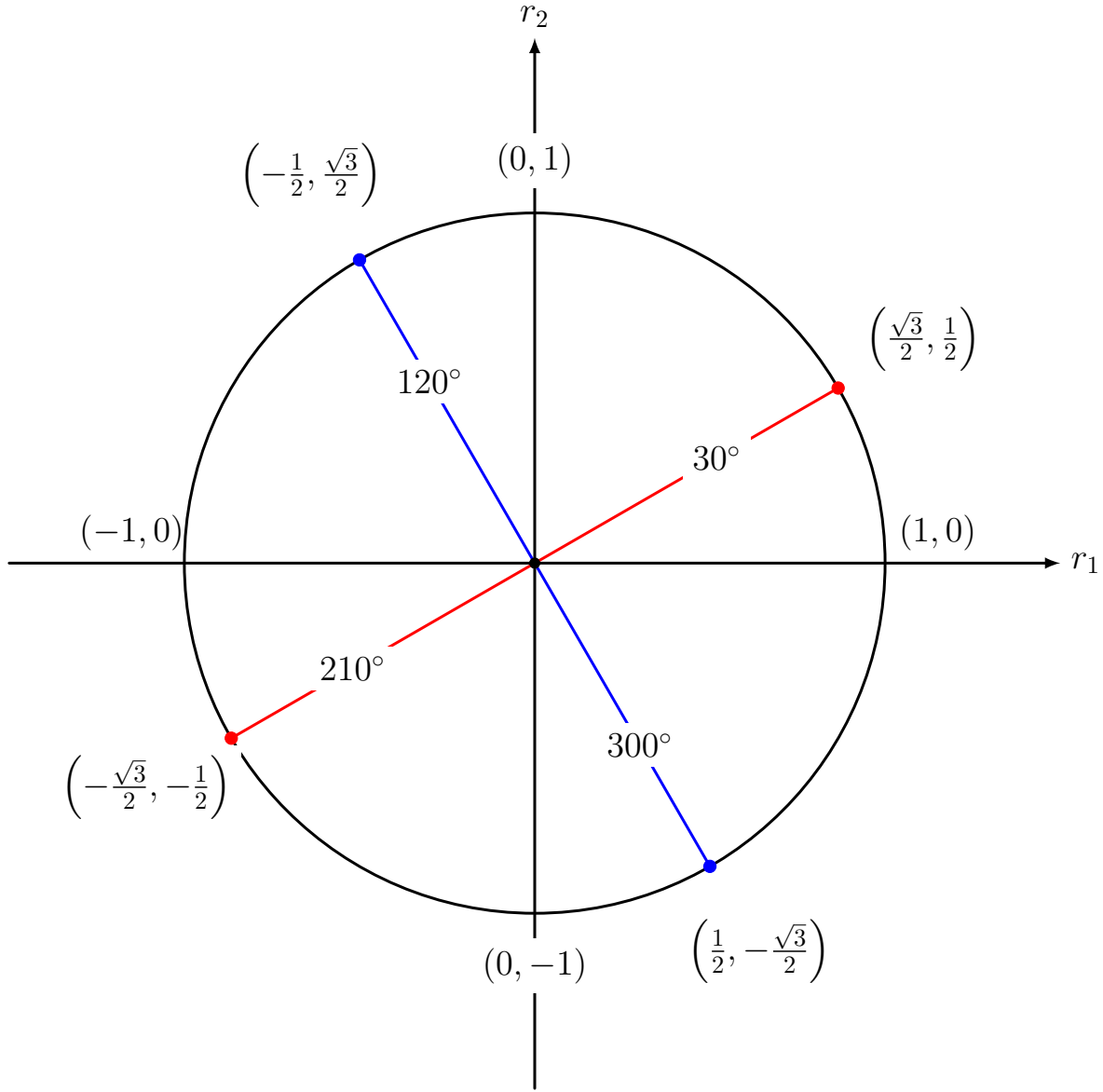

We can readily generalize this framework to the $N$-moduli case, in which
we define the radius vector $r$ to lie on the surface of an $\left( N-1 \right)$-sphere, and it is expressed as:
\begin{equation}
r = \left(r_1, r_2,..., r_N \right), \quad~\text{with}~\sum_{i = 1}^{N} r_i^2 = 1 \, ,
\label{radius3}
\end{equation}
while the antipodal vector $\bar{r}$ is given by:
\begin{equation}
\bar{r} = - \left(r_1, r_2,..., r_N \right) \, .
\label{radius4}
\end{equation}
As an illustration, we consider the three-field case: $N = 3$. In this case, the Minkowski solutions
lie anywhere on the surface of the unit sphere.
dS and AdS solutions can be obtained from any point on the sphere, by combining it with this antipodal point with $r_i \to -r_i$. 
In Fig.~\ref{sphere1} we show an example where four different Minkowski vacuum solutions 
are combined into 2 distinct Minkowski pairs lying on the surface of a sphere. 

\begin{figure}[h!]
\centering
\includegraphics[scale=0.9]{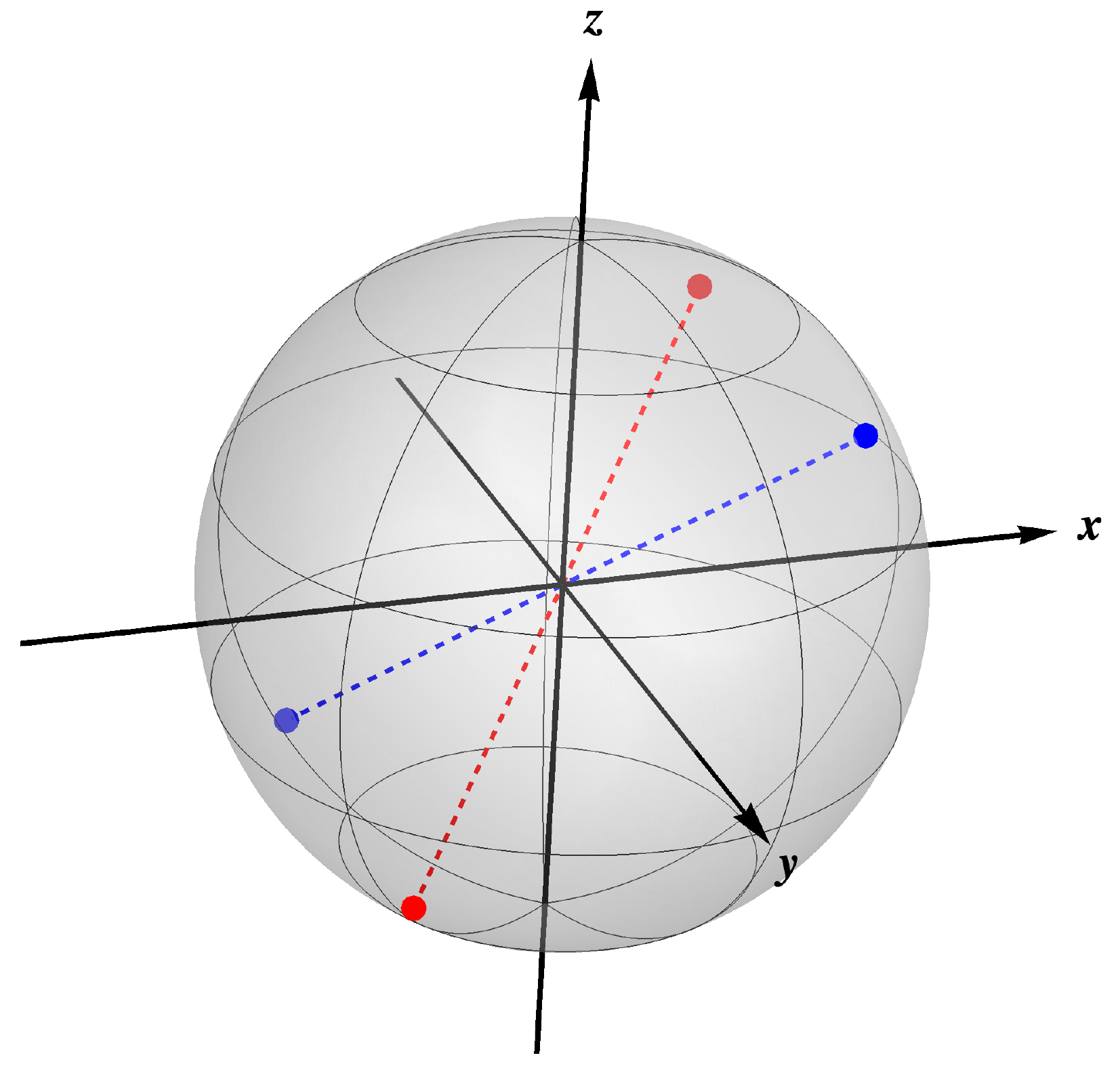}
\caption{\it Illustration of Minkowski pairs on the surface of a sphere. The sphere is split into eight octants 
and two distinct Minkowski pairs lying in different octants are shown. The red dots represent a 
Minkowski pair solution $r = (1/\sqrt{3}, - 1/\sqrt{3}, 1/\sqrt{3})$ and $\bar{r} = (-1/\sqrt{3},  1/\sqrt{3}, -1/\sqrt{3})$, 
which lies in the fourth and sixth octants of the sphere, while the blue dots represent a Minkowski pair solution 
$r =(1/\sqrt{3}, 1/\sqrt{3}, 1/\sqrt{3})$ and $\bar{r} = (-1/\sqrt{3}, -1/\sqrt{3}, -1/\sqrt{3})$, 
which lies in the first and seventh octants of the sphere.}
\label{sphere1}
\end{figure}

We have seen how all Minkowski pair solutions lie on the surface of an $\left(N-1 \right)$-sphere
of unit radius, and recall the general expressions for the corresponding powers, $n_{i}$ and  $\bar{n}_{i}$ of $\xi$ given earlier:
We show in Fig.~\ref{sheets} Minkowski pair solutions for these powers as functions of $|r_i|$ and $\alpha_i$.
The lower yellow sheet illustrates the possible choices for $n_{i}$, while the upper blue sheet illustrates the 
possible choices for $\bar{n}_{i}$. If we are only concerned with Minkowski solutions, 
we can freely choose any point lying on either the upper or lower sheet, which leads to $V = 0$. 
In order to construct successfully a Minkowski pair, we need to combine our chosen point with the
corresponding point on the opposite sheet, which will yield the dS/AdS solution $V = \, 12 \, \lambda_1 \, \lambda_2$.

\begin{figure}[h!]
\centering
\includegraphics[scale=.7]{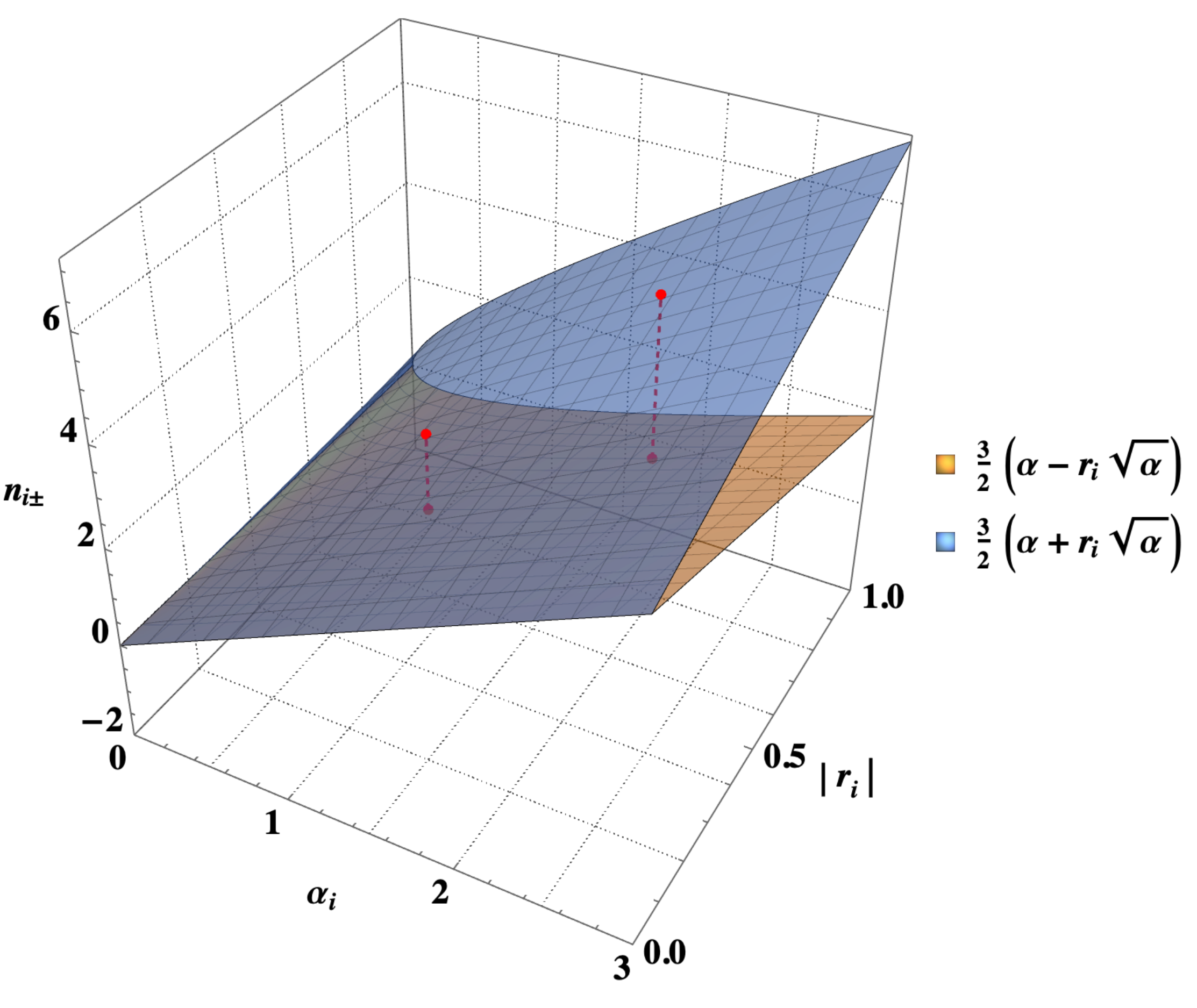}
\caption{\it Illustration of Minkowski pair formulation on the $n_{i}$ (yellow) and $\bar{n}_{i}$ (blue) sheets. The Minkowski pairs are depicted by red dots and their coordinates are given by $\left(1, \frac{1}{2}, \frac{3}{4} \right)$ with $\left( 1, \frac{1}{2}, \frac{9}{4} \right)$ and $\left( 2, \frac{3}{4}, 3 - \frac{9}{8} \sqrt{2} \right)$ with $\left( 2, \frac{3}{4}, 3 + \frac{9}{8} \sqrt{2} \right)$.}
\label{sheets}
\end{figure}

Having established successfully a geometric connection between unique vacuum solutions,
in the remaining sections we show that identical patterns emerge for K\"ahler potential forms 
with untwisted and twisted matter fields.

\section{Minkowski Pairs with Matter Fields}

\subsection{The Untwisted Case}
\label{untwisted}

In this Section, we extend  our formulation to no-scale models with untwisted matter fields. 
We begin by considering the following K\"ahler potential, which parametrizes a non-compact 
$\frac{SU(2,1)}{SU(2) \cross U(1)}$ coset space:
\begin{equation}
K \; = \; -3 \, \alpha \ln \left( T + \overline{T} - \frac{\phi \bar{\phi}}{3} \right) \, ,
\label{kah5}
\end{equation}
where $\alpha$ is a curvature parameter, $T$ can be interpreted as a volume modulus, and $\phi$ is a matter field.
Moreover, we impose the conditions $T = \overline{T}$ and 
$\phi = \bar{\phi}$ by fixing the VEVs of the imaginary components of the fields to zero, along the lines discussed above. 
Clearly Eq.~(\ref{kah5}) can be written in the form of Eq.~(\ref{kahv}) with $\mathcal{V}$ set equal to the
argument of the $\log$ in (\ref{kah5}), $\mathcal{V} = T + \overline{T} - \frac{\phi \bar{\phi}}{3}$.

Once again, when we restrict to real fields, and in this case set $T = \overline{T}$ and $\phi = \bar{\phi}$ we obtain:
\begin{equation}
\mathcal{V} \longrightarrow \xi, \quad~\text{with}~ \quad~\xi = 2T - \frac{\phi^2}{3} \, .
\end{equation}
We then consider the following form of superpotential:\\
\begin{equation}
W_{M} \; = \; \lambda \cdot \xi^n \, ,
\label{mink6}
\end{equation}
which leads to the following effective scalar potential:
\begin{equation}
V = \lambda^2 \, \xi^{2n - 3 \alpha} \cdot \left( \frac{(2n - 3 \alpha)^2}{3 \alpha} - 3 \right) \, .
\end{equation}
To obtain a Minkowski vacuum: $V = 0$, we solve the constraint for $n$, and recover the familiar result given in Eq.~(\ref{npm})
for the case with a single modulus. 
Using the superpotential in Eq.~(\ref{mink6}), 
we find the following scalar masses-squared for the imaginary components of the fields $T$ and $\phi$:
\begin{equation}
m_{Im\,T}^2 = \frac{4 \lambda^2 (\alpha - 1) \xi^{\pm 3 \sqrt{\alpha}}}{\alpha}, \quad m_{Im\,\phi}^2 = \frac{4 \lambda^2 (\sqrt{\alpha} \pm 1) \xi^{\pm 3 \sqrt{\alpha}}}{\sqrt{\alpha}} \, ,
\label{massim3}
\end{equation}
whereas, as anticipated, the masses-squared of the real components are $m_{Re\,T}^2 = 0$ and $m_{Re\,\phi}^2 = 0$.
It can be seen from Eqs.~(\ref{massim3}) that stability in the imaginary directions for both fields
requires that the inequalities $\alpha \geq 1$.}

To construct the $\frac{SU(2,1)}{SU(2) \cross U(1)}$ Minkowski pair formulation, we follow the previous discussion and 
use the same superpotential as in Eq.~(\ref{ds3})
\begin{equation}
W_{dS/AdS} = \lambda_1 \cdot \xi^{n_{-}} - \lambda_2 \cdot \xi^{n_{+}} \, .
\label{mink7}
\end{equation}
Doing so, we recover dS/AdS vacuum solutions given by Eq.~(\ref{potds}). 
In this case, the masses-squared for the imaginary field components are given by:
\begin{equation}
m_{Im\,T}^2 = \frac{ 4 \left(\lambda_1^2 (\alpha - 1) \xi^{-3 \sqrt{\alpha}} - 2 (\alpha + 1) \lambda_1 \lambda_2 + \lambda_2^2 (\alpha - 1) \xi^{3 \sqrt{\alpha}} \right)}{\alpha}
\label{massds4}
\end{equation}
and 
\begin{equation}
m_{Im\,\phi}^2 = 4 \left( \frac{\lambda_1^2 (\sqrt{\alpha} - 1) \xi^{-3 \sqrt{\alpha}}}{\sqrt{\alpha}} + 4 \lambda_1 \lambda_2 + 
 \frac{\lambda_2^2 (\sqrt{\alpha} + 1) \xi^{3 \sqrt{\alpha}}}{\sqrt{\alpha}}  \right) \, .
 \label{massds5}
\end{equation}
We do not discuss here the stabilization of these components, but we can always include quartic stabilization terms in 
the K\"ahler potential~(\ref{kah5}), as discussed previously.

Having established the principles in the case of the $\frac{SU(2,1)}{SU(2) \cross U(1)}$ K\"ahler potential with 
an untwisted matter field $\phi$, we can generalize our formulation to no-scale models that 
parametrize a non-compact $\frac{SU(N,1)}{SU(N) \cross U(1)}$ coset manifold. 
Following the same recipe considered in previous sections, we start with the K\"ahler potential (\ref{kahv}),
and we define the argument inside the logarithm as:
\begin{equation}
\mathcal{V} \equiv T + \overline{T} - \sum_{j = 1}^{N-1} \frac{|\phi_j|^2}{3} \, .
\end{equation}
Furthermore, we fix the VEVs of the imaginary fields to zero, so that $T = \overline{T}$ and $\phi_j = \bar{\phi}_j$. 
Using the same notation:
\begin{equation}
\mathcal{V} \longrightarrow \xi,~\text{when}~ T = \overline{T}~\text{and}~\phi_j= \bar{\phi}_j \, ,
\end{equation}
the argument inside the logarithm in the K\"ahler potential becomes
\begin{equation}
\xi = 2T - \sum_{j = 1}^{N-1} \frac{|\phi_j|^2}{3} \, .
\end{equation}
With this definition of $\xi$,  Minkowski vacuum solutions are found for the same choice of superpotential given in Eq.~(\ref{mink6}).
The masses-squared of the imaginary components, 
with $m_{Im\,T}^2$ and $m_{Im\,\phi_j}^2$ are given by~(\ref{massim3}).

At this point, it should not be surprising that by combining two distinct Minkowski solutions we can form a Minkowski superpotential pair given by 
Eq.~(\ref{mink7}).
This dS/AdS superpotential yields identical scalar masses-squared for the imaginary components, 
with $m_{Im\,T}^2$ given by~(\ref{massds4}) and $m_{Im\,\phi_j}^2$ given by~(\ref{massds5}). 

Finally, we can also extend our formulation to more complicated K\"ahler potentials that take the form 
$K = \sum_{i} K_i$, where each $K_i$ is of no-scale type and given by:
\begin{equation}
K = -3 \prod_{i = 1}^M \alpha_i \ln(\mathcal{V}_i), \quad \text{with}~\mathcal{V}_i =T_i + \overline{T}_i - \sum_{j = 1}^{N - 1} \frac{|\phi_{ij}|^2}{3} \, .
\end{equation}
We again assume that $T_i = \bar{T}_i$ and $\phi_{ij} = \bar{\phi}_{ij}$, which leads to:
\begin{equation}
\mathcal{V}_i \longrightarrow \xi_i,~\text{when}~T_i = \bar{T}_i~\text{and}~\phi_{ij} = \bar{\phi}_{ij} \, .
\end{equation}
Thus, we obtain the following Minkowski pair superpotential:
\begin{equation}
W_{dS/AdS} = \lambda_1 \cdot \prod_{i = 1}^M \xi_i^{n_{i}} - \lambda_2 \cdot \prod_{i = 1}^M \xi_i^{\bar{n}_{i}}, \quad \text{with}~V = 12 \, \lambda_1 \, \lambda_2\, ,
\end{equation}
which coincides with the multi-moduli case considered previously.

\subsection{The Twisted Case}
\label{twisted}

An analogous Minkowski pair formulation can also be considered in the case of twisted matter fields. 
We consider the corresponding K\"ahler potential:
\begin{equation} 
\label{s5e1}
K\; = \; -3 \, \alpha \, \ln \left( T + \overline{T} \right) + \varphi \bar{\varphi} \, ,
\end{equation}
where we introduce the notation $\varphi$ for twisted matter fields. To this end, we first find a relatively simple superpotential  form
that yields Minkowski solutions, and consider the following Ansatz:
\begin{equation}
\label{s5e2}
W_{M} = \lambda \cdot (2T)^n \cdot e^{-\varphi^2/2} \, .
\end{equation}
Combining it with the effective scalar potential in Eq.~(\ref{effpot}), and setting $T = \overline{T}$ and $\varphi = \bar{\varphi}$ by fixing the VEVs of the imaginary components of the fields to zero, we obtain:
\begin{equation}
V = \lambda^2 \cdot \left( 2T \right)^{2n - 3 \alpha} \cdot \left( \frac{(2n - 3 \alpha)^2}{3 \alpha} - 3 \right) \, .
\label{twistpot}
\end{equation}
From  the form of the scalar potential, we see that it does not depend on ${Re\,\varphi}$. To obtain a Minkowski vacuum solution, 
we find the same solutions found in Eq.~(\ref{npm}) for $n$. 
This yields the following scalar masses-squared for the imaginary components:
\begin{equation}
m^2_{Im\,T} \; = \; \frac{4\lambda^2 (\alpha-1)}{\alpha}\cdot (2T)^{\pm3\sqrt{\alpha}}
\label{mtws1}
\end{equation}
and
\begin{equation}
m^2_{Im\,\varphi} \; = \; 4\, \lambda^2 \, (2T)^{\pm3\sqrt{\alpha}} \, .
\label{mtws2}
\end{equation}
We can see from Eqs.~(\ref{mtws1}) and~(\ref{mtws2}) that ${Im\,\varphi}$ is always stable, 
and that ${Im\,T}$ is stable when $\alpha \geq 1$.

Similarly, we also consider the following Ansatz:
\begin{equation}
\label{s5e22}
W_{M} = \lambda \cdot (2T)^n \cdot e^{+\varphi^2/2} \, .
\end{equation}
If we combine this with Eq.~(\ref{effpot}), and set $T = \overline{T}$ and $\varphi = -\bar{\varphi}$, we obtain the same effective scalar potential (\ref{twistpot}) with solutions for $n$ given by Eq.~(\ref{npm}). In this case, the scalar potential does not depend on ${Im\,\varphi}$, and the scalar masses-squared are given by Eqs.~(\ref{mtws1}) and~(\ref{mtws2})~\footnote{It is important to note that in this case the effective scalar potential has curvature in the real direction and the scalar mass-squared expression~(\ref{mtws2}) becomes $m_{Re \,\varphi}^2$.}. Therefore, there are two ways to construct Minkowski vacuum solutions with twisted matter fields that do not depend on either the real or imaginary components of $\varphi$.

Next, we construct the dS/AdS superpotential by combining two distinct Minkowski solutions:
\begin{equation} 
\label{s5e3}
W_{dS/AdS} = \big( \lambda_1\cdot(2T)^{n_-}-\lambda_2\cdot(2T)^{n_+} \big)\cdot e^{-\varphi^2/2} \, ,
\end{equation}
where we choose a Minkowski pair construction which does not depend on ${Re\,\varphi}$, and, if we assume that $T = \overline{T}$ and $\varphi = \bar{\varphi}$, the effective scalar potential~(\ref{effpot}) 
is given by Eq.~(\ref{potds}) once again.
In the case of the superpotential~(\ref{s5e3}), the scalar masses-squared of the imaginary field components are:
\begin{equation} 
\label{s5e41}
m_{Im\,T}^2 \; = \; \frac{ 4 \left(\lambda_1^2 (\alpha - 1) \left(2T \right)^{-3 \sqrt{\alpha}} - 2 (\alpha + 1) \lambda_1 \lambda_2 + \lambda_2^2 (\alpha - 1) \left(2T  \right)^{3 \sqrt{\alpha}} \right)}{\alpha}
\end{equation}
and
\begin{equation}
m_{Im\,\varphi}^2 \; = \; 4 \left( \lambda_1^2 (2T)^{-3\sqrt{\alpha}} + 4 \lambda_1 \, \lambda_2 + \lambda_2^2 (2T)^{3\sqrt{\alpha}} \right) \, .
\label{s5e42}
\end{equation}
It is important to note that for de Sitter solutions, while $m_{Im\,\varphi}^2$ is always positive, 
$m_{Im\,T}^2$ is not and may require quartic stabilization terms in the
imaginary direction for the field $T$.

Analogously, one can also consider the following dS/AdS superpotential form:
\begin{equation}
W_{dS/AdS} = \big( \lambda_1\cdot(2T)^{n_-}-\lambda_2\cdot(2T)^{n_+} \big)\cdot e^{+\varphi^2/2} \, ,
\label{twist2}
\end{equation}
where, after setting $T = \overline{T}$ and $\phi = -\bar{\phi}$, we obtain the dS/AdS scalar potential $V = 12 \, \lambda_1 \lambda_2$, with the scalar masses-squared given by~(\ref{s5e41}) and~(\ref{s5e42}).

This analysis with a single twisted matter field can be generalized to include multiple fields. We consider the following K\"ahler potential form:
\begin{equation}
K \; = \; - 3 \, \alpha \, \ln (\mathcal{V}) + \sum_{j=1}^{N}|\varphi_j^2| \, .
\label{kahtw1}
\end{equation}
In this case, all of the previous results hold after the simple substitution of 
$\varphi^2 \to \sum \varphi_j^2$.

Another possible generalization is to consider K\"ahler potentials of the form $K = \sum_i K_i + \sum_j |\varphi_j|^2$, 
where each $K_i$ is of no-scale type:
\begin{equation}
K \; = \; - \, 3 \sum_{i = 1}^M \alpha_i \ln(\mathcal{V}_i)+\sum_{j=1}^N |\varphi_j|^2, \quad \text{with}~ \mathcal{V}_i = T_i + \overline{T}_i \, .
\label{kahtw2}
\end{equation}
As before, we assume that:
\begin{equation}
\mathcal{V}_i \longrightarrow \xi_i,~\text{when}~T_i = \overline{T}_i.
\end{equation}
In this case, with a superpotential of the form
\begin{equation}
W_{M} = \lambda \cdot \prod_{i = 1}^M \xi_i^{n_{i}}\cdot \exp \left(- \, \frac{1}{2}\sum_{j=1}^{N} \omega_j \varphi_j^2 \right),
\label{minktw1}
\end{equation}
where $\omega_j$ can take a value of either $+1$ or $-1$, we get a Minkowski solution $V=0$ after setting $T_i=\overline{T}_i$ and $\varphi_j = \omega_j \overline{\varphi}_j$. Similarly, we can obtain dS/AdS solutions $V= 12 \, \lambda_1 \, \lambda_2$ along the direction $T_i=\overline{T}_i$, $\varphi_j = \omega_j \overline{\varphi}_j$ from the superpotential:
\begin{equation}
W_{dS/AdS} = \left( \lambda_1 \cdot \prod_{i = 1}^M \xi_i^{n_{i}} - \lambda_2 \cdot \prod_{i = 1}^M \xi_i^{\bar{n}_{i}} \right)
\cdot \exp \left(- \, \frac{1}{2}\sum_{j=1}^{N} \omega_j\varphi_j^2 \right).
\label{dstw1}
\end{equation}

\subsection{The Combined Case}

We note finally that one can consider more complicated cases combining twisted and untwisted matter fields 
by following the principles discussed earlier in this Section. The only difference is that one needs to modify the K\"ahler potential in Eq.~(\ref{kahtw2}) and introduce untwisted matter fields $\phi_{ik}$:
\begin{equation}
K \; = \; - \, 3 \sum_{i = 1}^M \alpha_i \ln(\mathcal{V}_i)+\sum_{j=1}^N |\varphi_j|^2, \quad \text{with}~ \mathcal{V}_i = T_i + \overline{T}_i + \sum_{k = 1}^{P - 1} |\phi_{ik}|^2 \, ,
\end{equation}
If we assume that all our fields are fixed to be real, this leads to:
\begin{equation}
\mathcal{V}_i \longrightarrow \xi_i,~\text{when}~ T_i = \overline{T}_i,~\phi_{ik} = \bar{\phi}_{ik},~\text{and}~\varphi_j = \bar{\varphi}_j \, ,
\end{equation}
and for this case Minkowski solutions are given by superpotential~(\ref{minktw1}) and dS/AdS solutions are given by~(\ref{dstw1}).

\section{Applications to Inflationary Models}
\label{inflation}
\subsection{Inflation with an Untwisted Matter Field}

We now indicate briefly how to construct inflationary models in this framework \cite{ENOV2,ENOV3}.
For simplicity, we use a non-compact $\frac{SU(2, 1)}{SU(2) \cross U(1)}$ K\"ahler potential~(\ref{kah5}),
and we associate the matter field $\phi$ with the inflaton. If we set $\alpha = 1$, 
the K\"ahler potential~(\ref{kah5}) becomes:
\begin{equation}
K \; = \; -3 \ln \left(T + \overline{T} - \frac{\phi \bar{\phi}}{3} \right) \, .
\end{equation}
Next, we introduce a unified superpotential that combines the Minkowski pair superpotential $W_{dS}$ with an inflationary superpotential $W_{I} = f(\phi)$:
\begin{equation}
W = W_I + W_{dS} = M f(\phi) + \lambda_1 - \lambda_2 \left(2T - \frac{\phi^2}{3} \right)^3 \, .
\label{supunt}
\end{equation}
We also require that supersymmetry is broken at the minimum through the Minkowski pair superpotential $W_{dS}$ 
instead of the inflationary superpotential $W_{I}$. Therefore, we impose the conditions that $f(0) = f'(0) = 0$. 
Again, we assume that $T = \overline{T}$ and $\phi = \overline{\phi}$, and the superpotential~(\ref{supunt}) 
then yields the following effective scalar potential:
\begin{equation}
V \; = \; 12 \lambda_1 \, \lambda_2 + 12 \lambda_2 M f(\phi) + \frac{f'(\phi)^2}{\left(2T - \frac{\phi^2}{3} \right)^2} \, ,
\end{equation}
where we can safely neglect the mixing terms between $\lambda_2$ and $M$, leading to the approximation:
\begin{equation}
V \; \approx \; 12 \lambda_1 \, \lambda_2  + \frac{M^2 f'(\phi)^2}{\left(2T - \frac{\phi^2}{3} \right)^2} \, .
\end{equation}
Supersymmetry is broken by an $F$-term, which is given by:
\begin{equation}
\sum_{i = 1}^2 |F_i|^2 = F_T^2 = \left(\lambda_1 + \lambda_2 \right)^2,~\text{with}~m_{3/2} = \lambda_1 - \lambda_2 \, .
\label{fterm}
\end{equation}

\subsection{Inflation with a Twisted Matter Field}

Following the same approach, we now show how to construct viable inflationary models with a twisted inflaton field. 
We use a non-compact $\frac{SU(1, 1)}{U(1)} \cross U(1)$ K\"ahler potential form~(\ref{s5e1}),
and we associate the matter field $\varphi$ with the inflaton. We set $\alpha = 1$, and Eq.~(\ref{s5e1}) reduces to:
\begin{equation}
K \; = \; -3 \ln \left(T + \overline{T} \right) + \varphi \overline{\varphi} \, .
\end{equation}
Next, we introduce the following unified superpotential form
\footnote{Similarly, we can consider a unified superpotential form with $W_{dS}$ given by~(\ref{twist2}). In this case, inflation will be driven by ${Im\,\varphi}$.}:
\begin{equation}
W = W_I + W_{dS} = \bigg(M \, f(\varphi) + \lambda_1 - \lambda_2 \left(2T \right)^3 \bigg) \cdot e^{-\varphi^2/2} \, ,
\label{suptw}
\end{equation}
where the inflationary superpotential is given by $W_I = M \, f(\varphi) \cdot e^{- \varphi^2/2}$. 
We again require supersymmetry to be broken through the Minkowski pair superpotential $W_{dS}$, 
and we impose the conditions that at the minimum we must have $f(0) = f'(0) = 0$.
The superpotential form~(\ref{suptw}) leads to the following effective scalar potential:
\begin{equation}
V = 12 \lambda_1 \lambda_2 + 12 \lambda_2 M f(\varphi) + \frac{M^2 f'(\varphi)^2}{8T^3} \, .
\end{equation}
If we neglect the mixing terms between $\lambda_2$ and $M$, and fix $\langle T \rangle = \frac{1}{2}$, we can approximate:
\begin{equation}
V \approx 12 \lambda_1 \lambda_2  + M^2 f'(\varphi)^2 \, ,
\label{pottw2}
\end{equation}
and supersymmetry breaking is characterized by the same expression given in Eq.~(\ref{fterm}).
In order to construct a Starobinsky-like inflationary potential that is a function of the field $\varphi$,
we use the following canonical field redefinition:
\begin{equation}
\varphi = \frac{x + i y}{\sqrt{2}} \, ,
\end{equation}
and we assume that $\varphi = \overline{\varphi} = \frac{x}{\sqrt{2}}$. 
We then introduce the following inflationary superpotential form:
\begin{equation}
W_I =  \frac{3}{4}M \left(\frac{2 \varphi }{\sqrt{3}}+e^{-\frac{2 \varphi }{\sqrt{3}}}-1\right) e^{-\varphi^2/2} \, ,
\end{equation}
and assume that $\varphi = \overline{\varphi} = \frac{x}{\sqrt{2}}$ and $T = \overline{T} = 1/2$, which yields the Starobinsky inflationary potential with a positive cosmological constant at the minimum:
\begin{equation}
V = 12 \lambda_1 \lambda_2 + 3 \lambda _2 M \left(\sqrt{6} x+3 e^{-\sqrt{\frac{2}{3}} x}-3 \right) + \frac{3}{4} M^2 \left( 1 - e^{-\sqrt{\frac{2}{3}} x} \right)^2,
\end{equation}
or if we neglect the mixing terms between $\lambda_2$ and $M$, we obtain:
\begin{equation}
V \approx 12 \, \lambda_1 \lambda_2 + \frac{3}{4} M^2 \left( 1 - e^{-\sqrt{\frac{2}{3}} x} \right)^2.
\end{equation}

\section{Summary}
\label{summary}

We have exhibited in this paper the unique choice of superpotential leading to a Minkowski vacuum in a
single-field no-scale supergravity model, and also shown how to construct dS/AdS solutions 
using pairs of these single-field Minkowski superpotentials.
We have then extended these constructions to two- and multifield no-scale supergravity models, providing also a geometrical interpretation of the dS/AdS solutions in terms of combinations of superpotentials that are
functions of fields at antipodal points on hyperspheres. As we have also shown, these constructions can be
extended to scenarios with additional twisted or untwisted fields, and we have also discussed how
Starobinsky-like inflationary models can be constructed in this framework.

The models described in this paper provide a general framework that is suitable for
constructing unified supergravity cosmological models that include a primordial
near-dS inflationary epoch that is consistent with CMB measurements, the transition
to a low-energy effective theory incorporating soft supersymmetry breaking at some
scale below that of inflation, and a
small present-day cosmological constant (dark energy). As such, this framework
is suitable for constructing complete models of cosmology and particle physics
below the Planck scale.

For the future, two general classes of issues stand out. One is the construction of specific
models for sub-Planckian physics, which should address the incorporation of Standard Model
(and possibly other) matter and Higgs degrees of freedom. Should these be described by twisted or
untwisted fields, and how are they coupled to the inflaton? Specific answers to some of these
issues have been proposed in~\cite{EGNNO4}, and more details are forthcoming~\cite{EGNNO5}.

Another set of issues concerns the interface with string theory. For example, although no-scale supergravity
theories arise generically in the low-energy limits of string compactifications, many different
non-compact coset manifolds may be realized. Which of these is to be preferred? Another set of questions
concerns the specific forms of superpotential that are needed to obtain a Minkowski or dS vacuum. In this
paper we have constructed them from a bottom-up approach, and demonstrated their uniqueness.
How could one hope to obtain them in a top-down approach, starting from a specific string model?

This question is particularly acute in the case of dS vacuum solutions, since swampland conjectures~\cite{Swamp}
suggest that string theory may not possess such vacua. At the time of writing controversy still swirls
about these conjectures, and in this paper we have taken the pragmatic approach of exploring what
such solutions would look like. As such, our solutions may suggest avenues to explore in searching
for them, or at least the obstacles to be overcome. The existence or otherwise of dS vacua in string theory is
clearly a key issue for the future that lies beyond the scope of this paper.

\subsection*{Acknowledgements}

\noindent
The work of JE was supported in part by the United Kingdom STFC Grant
ST/P000258/1, and in part by the Estonian Research Council via a
Mobilitas Pluss grant. The work of DVN was supported in part by the DOE
grant DE-FG02-13ER42020 and in part by the Alexander~S.~Onassis Public
Benefit Foundation. The work of KAO was
supported in part by DOE grant DE-SC0011842 at the University of
Minnesota.

\end{document}